\documentclass{emulateapj}
\usepackage{times,amssymb,hyperref,subfigure,epsfig,aas_macros}

\newcommand{\zp}{\texttt{Zodipic}}

\begin{document}


\title{Exo-zodi modelling for the Large Binocular Telescope Interferometer}
\shorttitle{Exo-zodi modelling for the LBTI}
\shortauthors{G. M. Kennedy et al.}

\author{Grant M. Kennedy$^1$, Mark C. Wyatt$^1$, Vanessa Bailey$^2$, Geoffrey Bryden$^3$,
  William C. Danchi$^4$, Denis Defr\`ere$^2$, Chris Haniff$^5$, Philip M.  Hinz$^2$,
  J\'er\'emy Lebreton$^{6,7}$, Bertrand Mennesson$^3$, Rafael Millan-Gabet$^7$, Farisa
  Morales$^3$, Olja Pani\'c$^1$, George H. Rieke$^2$, Aki Roberge$^4$, Eugene
  Serabyn$^3$, Andrew Shannon$^1$, Andrew J. Skemer$^2$, Karl R. Stapelfeldt$^4$,
  Katherine Y. L. Su$^2$, Alycia J. Weinberger$^8$}

\affil{$^1$Institute of Astronomy, University of Cambridge, Madingley Road, Cambridge CB3 0HA,
  UK \\
  $^2$Steward Observatory, University of Arizona, 933 North Cherry Avenue, Tucson, AZ
  85721, USA \\
  $^3$Jet Propulsion Laboratory, California Institute of Technology, 4800 Oak Grove
  Drive, Pasadena, CA 91109, USA \\
  $^4$NASA Goddard Space Flight Center, Exoplanets and Stellar Astrophysics, Code 667,
  Greenbelt, MD 20771, USA \\
  $^5$Cavendish Laboratory, University of Cambridge, JJ Thomson Avenue, Cambridge CB3 0HE, UK \\
  $^6$Infrared Processing and Analysis Center, MS 100-22, California Institute of
  Technology, 770 S. Wilson Ave., Pasadena, CA 91125, USA \\
  $^7$NASA Exoplanet Science Institute, California Institute of Technology, 770 S. Wilson
  Ave., Pasadena, CA 91125, USA \\
  $^8$Department of Terrestrial Magnetism, Carnegie Institution of Washington, 5241 Broad
  Branch Road
  NW, Washington, DC 20015, USA \\
}

\begin{abstract}
  Habitable zone dust levels are a key unknown that must be understood to ensure the
  success of future space missions to image Earth analogues around nearby stars. Current
  detection limits are several orders of magnitude above the level of the Solar System's
  Zodiacal cloud, so characterisation of the brightness distribution of exo-zodi down to
  much fainter levels is needed. To this end, the Large Binocular Telescope
  Interferometer (LBTI) will detect thermal emission from habitable zone exo-zodi a few
  times brighter than Solar System levels. Here we present a modelling framework for
  interpreting LBTI observations, which yields dust levels from detections and upper
  limits that are then converted into predictions and upper limits for the scattered
  light surface brightness. We apply this model to the HOSTS survey sample of nearby
  stars; assuming a null depth uncertainty of 10$^{-4}$ the LBTI will be sensitive to
  dust a few times above the Solar System level around Sun-like stars, and to even lower
  dust levels for more massive stars.
\end{abstract}


\section{Introduction}\label{s:intro}

One of the major long-term goals of astronomy is to place the Solar System within a
greater context, finding for example whether habitable planets like the Earth are
typical, and whether these Earth analogues have conditions suitable for alien life. Such
goals are ambitious and many obstacles must be overcome for them to come to fruition.

For the particular case of directly imaging Earth-like planets around other stars, a
major unknown is the level of photon noise from the dust that resides in the target
system itself, specifically dust located in the ``habitable zone'' \citep[HZ,
e.g.][]{1993Icar..101..108K}. Such dust populations are generically referred to as
``exo-zodi'' by analogy with the Solar System's Zodiacal cloud, though they may have
different origins. To be clear, we consider the dust populations seen around 10-25\% of
nearby stars with near-infrared (IR) interferometry to be a largely unrelated phenomenon,
based on their small grain size and hot temperatures
\citep[e.g.][]{2011A&A...534A...5D,2013A&A...555A.146L}, and lack of correlation with
both mid-IR detections of HZ dust (Mennesson et al., ApJ, in press) and far-IR detections
of cool dust \citep{2013A&A...555A.104A,2014arXiv1409.6143E}.

If HZ dust levels an order of magnitude or so greater than the Solar System level are
typical, this noise source could seriously hinder an Earth analogue-imaging and
characterisation mission \citep[which we will refer to as ``Earth-imaging'' for brevity,
e.g.][]{2010ASPC..430..293A,2012PASP..124..799R,2014arXiv1409.5128S,2014arXiv1402.2612B}.
Currently, the brightness distribution (which we also call the luminosity function) of
exo-zodi is largely unknown. Limits have been set by the Keck Interferometer Nuller
\citep[KIN,][Mennesson et al., ApJ, in press]{2012ApJ...748...55S,2011ApJ...734...67M},
but the distribution has only been characterised at levels 3-4 orders of magnitude above
the Solar System by photometric methods \citep{2013MNRAS.433.2334K}.

Enter the Large Binocular Telescope Interferometer (LBTI), a mid-IR instrument
specifically designed to characterise exo-zodi at dust levels just a few times greater
than the Solar system's Zodiacal cloud. For a brief description of the instrument see
Defr\`ere et al. (submitted). By observing of order 50 nearby stars at this level of
sensitivity (Weinberger et al., submitted), the LBTI will identify specific targets with
low dust levels that are suitable for future Earth-imaging, and moreover will
characterise the exo-zodi luminosity function with sufficient detail to provide new
information on the dust origin and evolution. This latter point is important because the
LBTI will observe a limited number of stars and cannot access the entire sky, and
population-level information such as correlations with other system properties will be
needed to assess the suitability of targets that are not observed with the LBTI.

Of course the results will also be scientifically valuable and interesting, particularly
when combined with existing and future observations (e.g. Defr\`ere et
al. submitted). Mid and far-IR imaging using telescopes such as \emph{Spitzer, Herschel},
and JWST will provide important information on the existence, location, and structure of
cooler dust belts that lie outside the HZ. Interferometric observations with longer
baselines (e.g. VLTI/MATISSE) and/or at different wavelengths (e.g. CHARA/FLUOR,
VLTI/PIONIER) will provide information that complements the LBTI to help build a complete
picture of the inner regions of individual systems.

The goal of this paper is to outline a modelling framework for interpreting LBTI
observations. Such models are needed because LBTI observations yield limited information
on the spatial structure of any disk that is detected. It is therefore important that the
model used to interpret the observations is thoroughly characterised so that the
constraints placed on its various parameters can be understood in terms of their
implications for the disk structure and surface brightness. The primary goal is to use
the model to make a useful statement about the level of dust in a system, or the limits
on undetected dust, in the habitable zone where Earth-imaging will be attempted in the
future.

In what follows, we outline a parameterised dust model that can be used to approximate
the Solar System's zodiacal dust cloud. We show how this model can be used to derive the
distribution of dust levels given an LBTI observation of thermal dust emission. We then
determine the corresponding implications for scattered light levels at visible
wavelengths. We finally show the expected levels of dust that could be detected by the
Hunt for Observable Signatures of Terrestrial Planetary Systems (HOSTS) survey and the
corresponding scattered light levels. For a detailed description of the LBTI sample we
refer the reader to Weinberger et al. (submitted), and for discussion of the instrument
and the first scientific results to Defr\`ere et al. (submitted).

\section{Physical model}\label{s:model}

The primary goal of our exo-Zodiacal cloud model is to allow easy comparison of the
results among survey stars, which span a range of distances and luminosities. In general
we expect that detections will be near the sensitivity limits, and that there will be
little information about any warm dust detected (or not detected) other than from
LBTI. Therefore, while warm dust may arise from various different processes, for example
\emph{in situ} Asteroid-belt like evolution, stochastic collisions, delivery of comets
from elsewhere, or some combination of all three
\citep[e.g.][]{2007ApJ...658..569W,2010ApJ...713..816N,2012MNRAS.tmp.3462J,2013MNRAS.433.2334K},
we will have little power to distinguish among them. Some scenarios could result in
clumpy non-axisymmetric structures, but there again is little hope of distinguishing
these from axisymmetric structures in all but the dustiest systems where LBTI detections
will be relatively easy and at high signal to noise ratios. Therefore, the models
considered here are axisymmetric.

\subsection{The star}

For specific stars, the stellar luminosity $L_\star$ (in Solar units) is easily derived
by fitting stellar atmosphere models to photometry, and the flux density $F_{\nu,\star}$
at any wavelength can be inferred from these models. Given a distance $d$, the stellar
radius can be estimated from the same model, though in some cases it may have been
directly observed. In any case, the contribution of the resolved stellar disk to an LBTI
observation will generally be negligible. Here we use the stellar fluxes for HOSTS survey
stars derived by Weinberger et al. (submitted).

\subsection{The disk}\label{ss:model}

LBTI observations are sensitive to the surface brightness distribution $S_{\rm disk}$ in
the N' band at 11$\mu$m (the bandpass spans 9.81-12.41 $\mu$m). Our model can be
considered a means of parameterising this surface brightness distribution using
parameters that have some physical relevance. Unless stated otherwise, in what follows
our calculations are made at 11 $\mu$m.

The surface brightness profile of the disk, if viewed face-on, is modelled as
\begin{equation}\label{eq:sb}
  S_{\rm disk} = 2.35 \times 10^{-11} \Sigma_{\rm m} B_\nu(\lambda,T_{\rm BB}) .
\end{equation}
Here $T_{\rm BB}$ is the temperature of disk material that behaves like a black body
\begin{equation}
T_{\rm BB}(r) = 278.3 L_\star^{0.25} r^{-0.5} {\rm K},
\end{equation}
where the distance to the star, $r$, is in AU.  The numerical factor in eq.~\ref{eq:sb}
is a conversion (1AU$^2$/1pc$^2$) so that the surface brightness is in units of Jy
arcsec$^{-2}$. The parameter $\Sigma_{\rm m}$ is representative of the disk's face-on
surface density of cross-sectional area (in AU$^2$/AU$^2$), so is analogous to optical
depth and is assumed to have a power law distribution
\begin{equation}
  \Sigma_{\rm m}(r) = z \Sigma_{\rm m,0} (r/r_0)^{-\alpha}
\end{equation}
between radii $r_{\rm in}$ and $r_{\rm out}$ (both in AU).  The normalisation
$\Sigma_{\rm m,0}$ is to be set at some $r_0$ (in AU) such that the surface density is in
units of zodis $z$ (see section \ref{ss:zodi}).

For generality we include an inclination $I$, which is half the total opening angle of
the disk, and represents the maximum inclination of the parent bodies (whose nodes would
be assumed to be randomised). Here we assume that our disks have negligible opening
angles so $I$ is not used, but that $I$ is always sufficiently large that the disk is
radially optically thin.

If all of the disk particles behave like black bodies, the parameter $\Sigma_{\rm m}$ can
be interpreted as the disk's surface density of cross-sectional area. Its interpretation
becomes more complicated for disks with more realistic particle optical properties
because the observed dust can be several times hotter than the black body temperature,
and not all radiation incident on a particle is absorbed and thermally re-emitted.
However, as the surface brightness is typically assumed to exhibit a power law dependence
across the relevant radius range, we have not lost any generality by expressing the disk
surface brightness in this way, even if the disk's temperature profile differs from that
of a black body. These issues will be discussed in more detail in sections \ref{ss:other}
and \ref{ss:scat}.

The final parameters, which specify the direction to the observer relative to the disk,
are the disk inclination $i$ and the position angle $\Omega$. The inclination is defined
such that $i=0$ is face-on and the position angle is measured anticlockwise (i.e. East)
from North. The range of inclinations is 0 to $90^\circ$ and the range of position angles
0 to $180^\circ$,\footnote{The symmetry of the LBTI transmission pattern (described in
  section \ref{ss:lbtitrx}) means that if the disks are not spatially resolved we in fact
  only need to consider angles between 0 and 90$^\circ$. We retain the larger range
  because we may wish to set $\Omega$ based on the position angle of the star or an outer
  disk component.} with these ranges set because we cannot distinguish between the near
and far side of a disk. That is, the faint disks considered here are optically thin
viewed from any direction, and in thermal emission look the same mirrored in the sky
plane or for 180$^\circ$ rotations. The model parameters and their meanings are
summarised in Table \ref{tab:params}.

\begin{table}
  \caption{Model parameters}\label{tab:params}
  \begin{tabular}{llll}
    \hline
    symbol & unit & parameter & reference \\
    \hline
    $d$ & pc & Distance to star+disk system & - \\
    $L_\star$ & $L_\odot$ & Stellar luminosity & 1 \\
    $F_\star$ & Jy & Stellar flux density at wavelength $\lambda$ & - \\
    $R_\star$ & $R_\odot$ & Stellar radius & 1 \\
    \hline
    $r_{\rm in}$ & AU & Inner disk radius (1500K) & 0.034\\
    $r_{\rm out}$ & AU & Outer disk radius (88K) & 10 \\
    $r_0$ & AU & Reference radius ($=\sqrt{L_\star/L_\odot}$ AU) & 1 \\
    $\Sigma_{\rm m,0}$ & - & Surface density at $r_0$ for a $z=1$ disk & $7.12
    \times 10^{-8}$ \\
    $z$ & - & Surface density at $r_0$ in zodis & 1 \\
    $\alpha$ & - & Index for $\Sigma_{\rm m}$ & 0.34 \\
    $I$ & $^\circ$ & Disk half-opening angle & small \\
    \hline
    $i$ & $^\circ$ & Disk inclination relative to face-on & - \\
    $\Omega$ & $^\circ$ & Disk position angle E of N on the sky & - \\
    \hline
  \end{tabular}  
  \tablecomments{Model parameters are separated into those related to the star, disk, and 
    observation. The rightmost column gives parameters for the reference disk
    model.}
\end{table}

\subsubsection{\zp}\label{sss:zodipic}

\zp~is an implementation of the \citet{1998ApJ...508...44K} zodiacal cloud
model.\footnote{Freely available IDL software written by M. Kuchner} This parametric
model has many parameters, whose values were derived via fitting to COBE/DIRBE
observations. \zp~produces two-dimensional images of the Solar System's Zodiacal dust
cloud as might be seen by an external observer, and has been used as a reference model
for calculating the limits set by KIN observations \citep{2011ApJ...734...67M}. Though it
is a complex model and can include dust components such as the Earth's resonant ring and
trailing blob, only the main smooth axisymmetric component is normally used. Because this
main component is simply a radial power law, with an approximately Gaussian vertical
density distribution, it is possible to reproduce the \zp~surface brightness using our
parameterised model.

The vertical dust distribution is the main difference between the
\citet{1998ApJ...508...44K} model and ours. The \citet{1998ApJ...508...44K} model assumes
that the radial and vertical components of the cloud are separable
\begin{equation}\label{eq:zodipic}
  n = n_0 r^{-p} \exp\left(-\beta g^\gamma\right)
\end{equation}
where
\begin{equation}\label{eq:zodipic2}
  g = \left\{ \begin{array}{ll} 
      \xi^2/2\mu & \xi < \mu \\
     \xi - \mu/2  & \xi \ge \mu
    \end{array}
  \right.
\end{equation}
and $\xi = |Z/r|$ and $Z$ is the height above the disk midplane in AU. The best fit
values for the COBE data were $n_0=1.13 \times 10^{-7}$, $p=1.34$, $\beta=4.14$,
$\gamma=0.942$, and $\mu=0.189$. The units of the normalisation are AU$^2$/AU$^3$ (i.e. a
volume density of surface area), and the integrated length of the vertical term is not
constant with radius, so their exponent $p$ is not the same as our exponent
$\alpha$. Numerically integrating equation (\ref{eq:zodipic2}) over $-\infty < \xi <
\infty$ yields 0.63, so the surface density at 1 AU is $7.12 \times 10^{-8}$
(AU$^2$/AU$^2$). Assuming for mathematical simplicity that $\gamma=1$, the integral of
the exponential term in equation (\ref{eq:zodipic}) is $\propto r$ for both cases of $g$,
and hence $\alpha \approx p-1$ and $\alpha \approx 0.34$. Therefore, the \zp~model can be
reasonably expressed in terms of our model with $r_0=1$ AU $\Sigma_{\rm m,0} = 7.12
\times 10^{-8}$, and $\alpha = 0.34$.

There remain a few small differences in the surface brightness of our model relative to
\zp, as shown in the right panel of Fig. \ref{fig:mod}. The main difference is that the
\zp~model is slightly brighter at larger radii because the radial temperature profile is
flatter. Setting the \zp~temperature profile to equal our black body prescription leads to
nearly indistinguishable face-on surface brightness profiles. Some minor differences
arise because the \citet{1998ApJ...508...44K} model defines radius as
$r=\sqrt{x^2+y^2+z^2}$, whereas our narrow opening angle means that our model is
effectively cylindrical.

\subsubsection{Reference model}\label{sss:ref}

\begin{figure*}
  \begin{center}
    \hspace{-0.5cm} \includegraphics[width=0.32\textwidth]{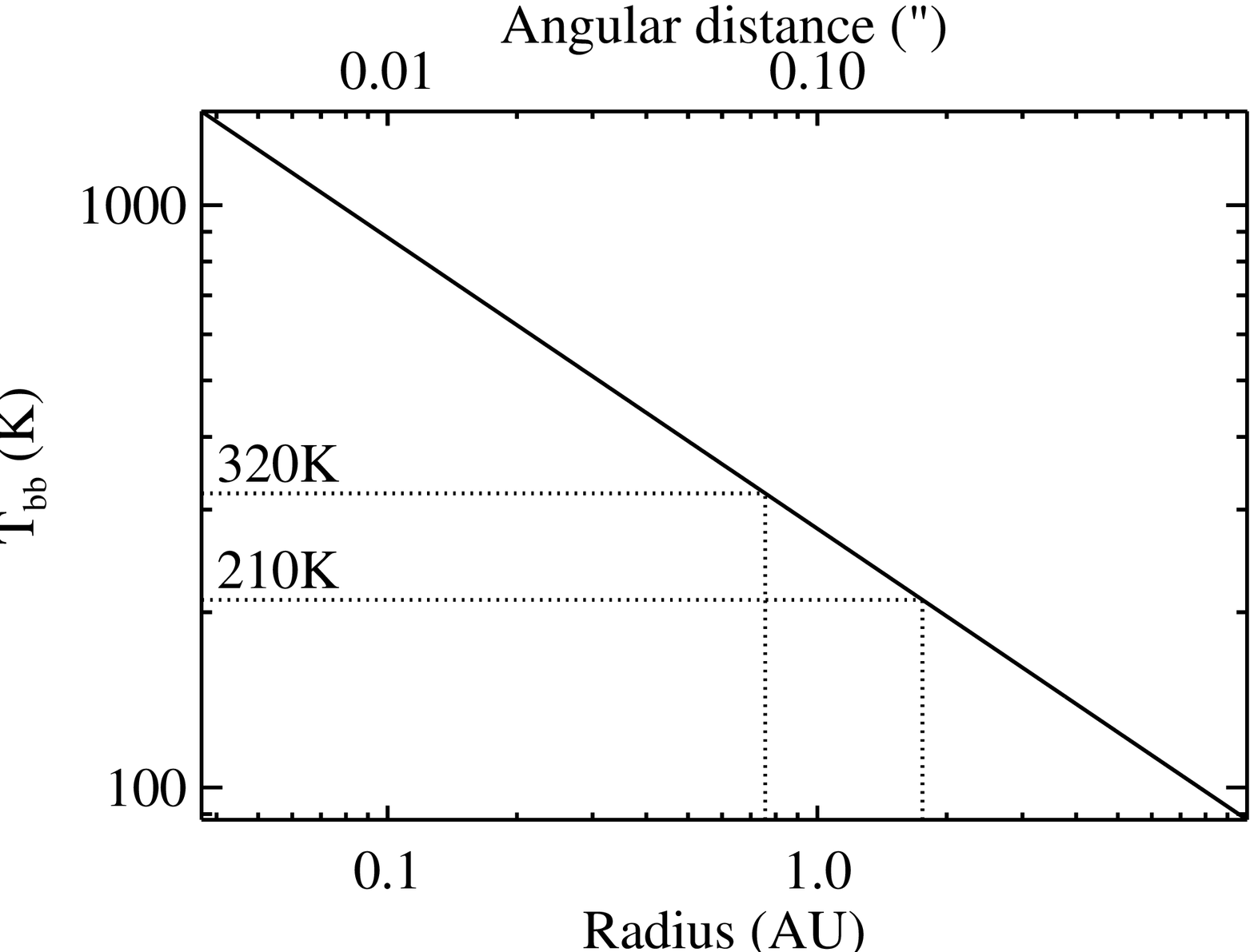}
    \hspace{0.2cm} \includegraphics[width=0.32\textwidth]{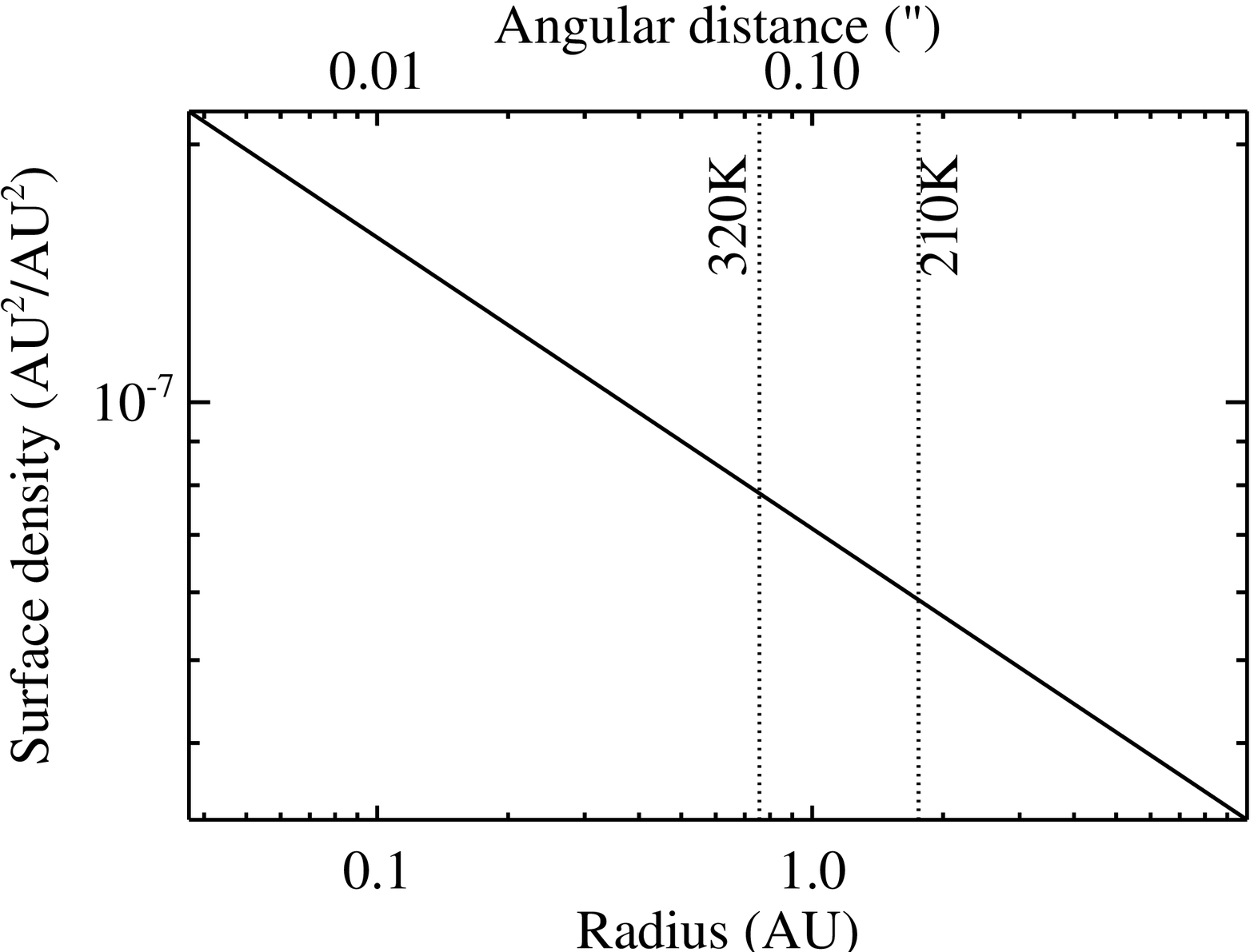}
    \hspace{0.2cm} \includegraphics[width=0.32\textwidth]{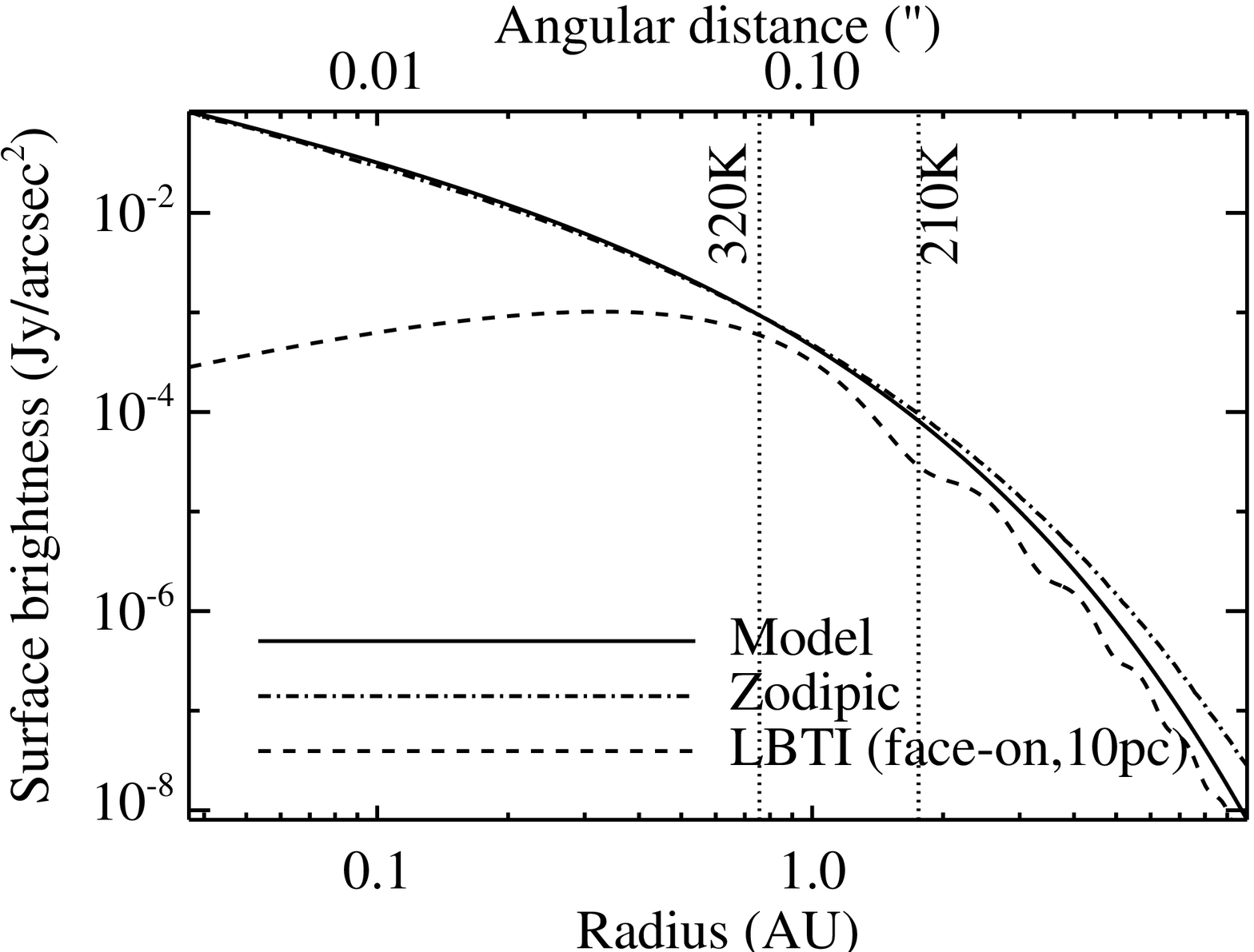}
    \caption{Reference disk model as described in section \ref{sss:ref}, with an
      approximate habitable zone between 210 and 320 K marked by dotted lines. The left
      panel shows the model temperature. The middle panel shows the disk surface density
      of cross sectional area. The right panel shows the disk surface brightness of our
      reference model (solid line) and an equivalent \zp~model for comparison (dot-dashed
      line), and the flux that would be transmitted through the LBTI for our reference
      model disk seen face-on at 10pc (dashed line).}\label{fig:mod}
  \end{center}
\end{figure*}

Here we introduce a reference disk model, the parameters of which are summarised in Table
\ref{tab:params} and based on the \citet{1998ApJ...508...44K} Solar System model
described above. The model extends from 0.034-10 AU with a power-law index $\alpha=0.34$,
and is scaled to $\Sigma_{\rm m,0}= 7.12 \times 10^{-8}$ at $r_0=1$ AU (the location
where $T_{\rm BB}=278.3$ K). The inner edge is set by a black body temperature of 1500K,
approximately the sublimation temperature of silicates, though we show in section
\ref{ss:other} that this choice is irrelevant for our model as long as the inner disk
edge is inside the IWA of 40 mas. The outer edge of 10 AU is chosen to be sufficiently
large that it does not affect the model at the LBTI wavelength of 11 $\mu$m. As noted
above we treat the disk as two dimensional so the opening angle is small. Our reference
model is based on the Solar System so the star is Sun-like, but as described in section
\ref{ss:zodi} the reference parameters are later varied to model disks around different
stars.

Figure \ref{fig:mod} shows the distribution of temperature, surface density and surface
brightness at 11$\mu$m in this reference model, where we have assumed a distance of 10
pc.  Survey targets are chosen so that the LBTI is sensitive to dust in their habitable
zone (Weinberger et al., submitted), so the dotted lines show the range of radii implied
by a temperature range of 210-320K, a rough indication of the habitable zone location for
the Sun \citep[e.g.][]{2013arXiv1301.6674K}. The right panel shows that the surface
brightness of our model closely matches \zp, and as noted above this difference only
arises due to different assumptions about the radial temperature profile. Therefore, for
the same assumptions and aside from the differences in scale height our model yields the
same results as \zp~for a Sun-like star.

\subsubsection{Normalisation (What is a zodi?)}\label{ss:zodi}

The disk normalisation (i.e. brightness) is set by the parameter $\Sigma_{\rm m,0}$,
which is the disk surface density at $r_0$. The goal here is to set what $\Sigma_{\rm
  m,0}$ and $r_0$ are so that they scale sensibly with stellar spectral type, and so that
our disk model is somehow sensibly related to zodi units, and therefore the Solar System
dust level. These parameters, along with $r_{\rm in}$, $r_{\rm out}$, $\alpha$, and $I$
define our reference disk model. In making this definition, it is important to remember
the ultimate goal of the LBTI, which is to constrain the level of dust in the habitable
zone around nearby stars. That is, disk surface brightness is the most important measure
of the impact of exo-zodi on future Earth-imaging missions.

A desirable definition might be one of ``constant hindrance'' for an Earth-imaging
mission, where $\Sigma_{\rm m,0}$ is set such that a $z=1$ disk impacts all observations
in search of exo-Earths at the same level. However, for this imaging any exo-Earths will
be unresolved, whereas the exo-zodi that accompany them will be resolved, so the same
physical disk has the same surface brightness regardless of distance, but the brightness
of the unresolved planet decreases with distance. Therefore, any zodi definition that
attempts to provide constant hindrance will be distance dependent, which is an
undesirable property since it is unrelated to the physical disk structure.

To set $\Sigma_{\rm m,0}$ therefore requires a choice between using either the total disk
brightness, or the surface brightness at $r_0$. Both approaches have their merits. For
example, using the total disk brightness allows comparison with photometric observations
and therefore easy construction of the warm dust luminosity function
\citep[e.g.][]{2013MNRAS.433.2334K}. However, as we show below in Fig. \ref{fig:fnutvsr},
LBTI observations are not necessarily sensitive to total disk brightness (e.g. when a
significant fraction of the disk emission lies inside the first transmission peak).

We therefore prefer a zodi definition linked to the disk surface density in the habitable
zone. For the spectral type scaling, it then makes sense for $r_0$ to scale with
$\sqrt{L_\star}$, so that the surface brightness expressed by $\Sigma_{\rm m,0}$
corresponds to the radial distance where the equilibrium temperature is the same as at
Earth (i.e. an ``Earth-equivalent'' distance), therefore $r_0 = \sqrt{L_\star/L_\odot}$
AU. Scaling $r_{\rm in}$ and $r_{\rm out}$ by $\sqrt{L_\star/L_\odot}$ is also
appropriate, since this results in a common temperature at the inner and outer edges of
the disk, and maintains the disk inner edge at the sublimation radius. The disk edges are
of relatively little importance here because as we show below the LBTI is relatively
insensitive to their location.

With this definition, $\Sigma_{\rm m,0}$ is fixed at the surface density of the Solar
System's zodiacal cloud at 1 AU, but the location where this surface density applies
depends on the stellar luminosity. This definition is different to \zp, which fixes $r_0$
at 1 AU for all stars and fixes $r_{\rm in}$ at the sublimation radius. Because the dust
temperature at $r_0$ is the same for all spectral types, the thermal surface brightness
at $r_0$ (i.e. in the habitable zone) for a 1 zodi disk is also the same. Because it is
based on surface brightness, this zodi definition is well suited for use as an important
metric for future Earth-imaging missions.

An LBTI observation is not necessarily sensitive to the radial extent of a disk, so the
derived zodi level depends on what is assumed for $r_{\rm in}$ and $r_{\rm out}$,
particularly if the disk is relatively narrow. For example, below we consider a ``worst
case'' scenario where the dust emission only originates in the habitable zone, in which
case the derived dust surface density (and hence $z$) is higher for a given null
depth. To avoid confusion we recommend that zodi levels primarily use our reference model
to allow consistent comparisons.

\subsection{Transmission through the LBTI fringe pattern}\label{ss:lbtitrx}

\begin{figure*}
  \begin{center}
    \hspace{-0.15cm} \includegraphics[width=0.33\textwidth]{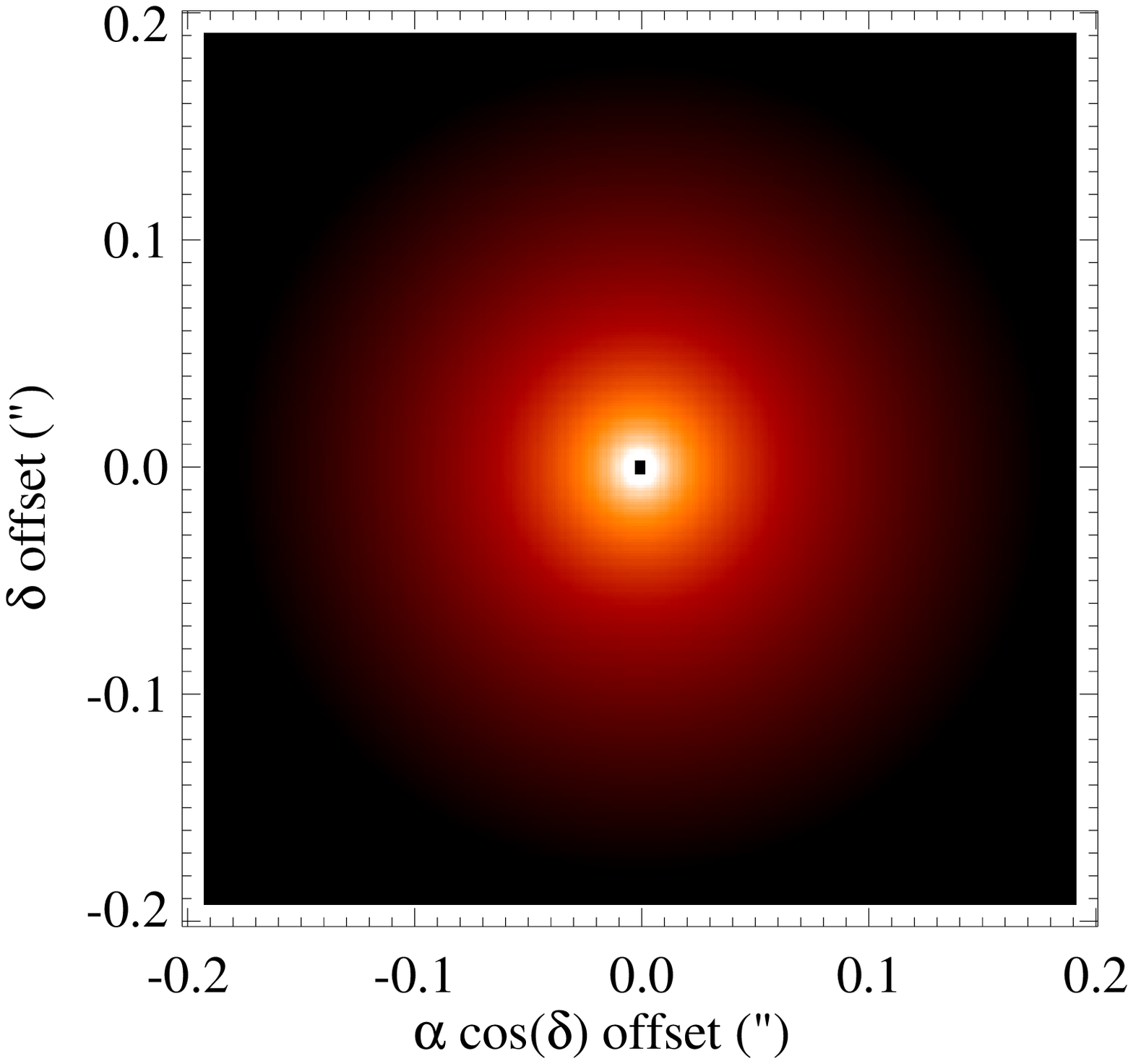}
    \hspace{-1cm} \includegraphics[width=0.33\textwidth]{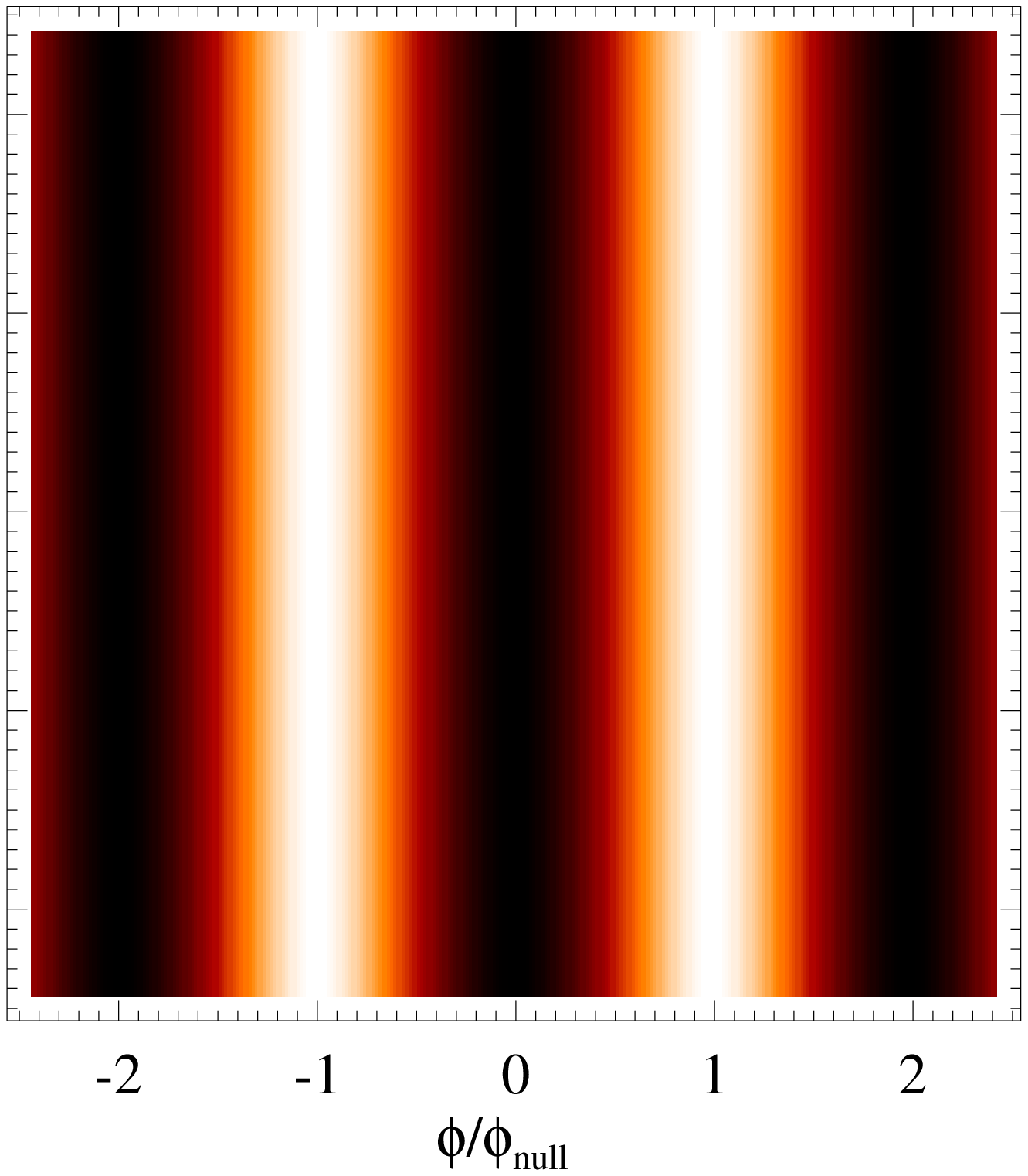}
    \hspace{-1cm} \includegraphics[width=0.33\textwidth]{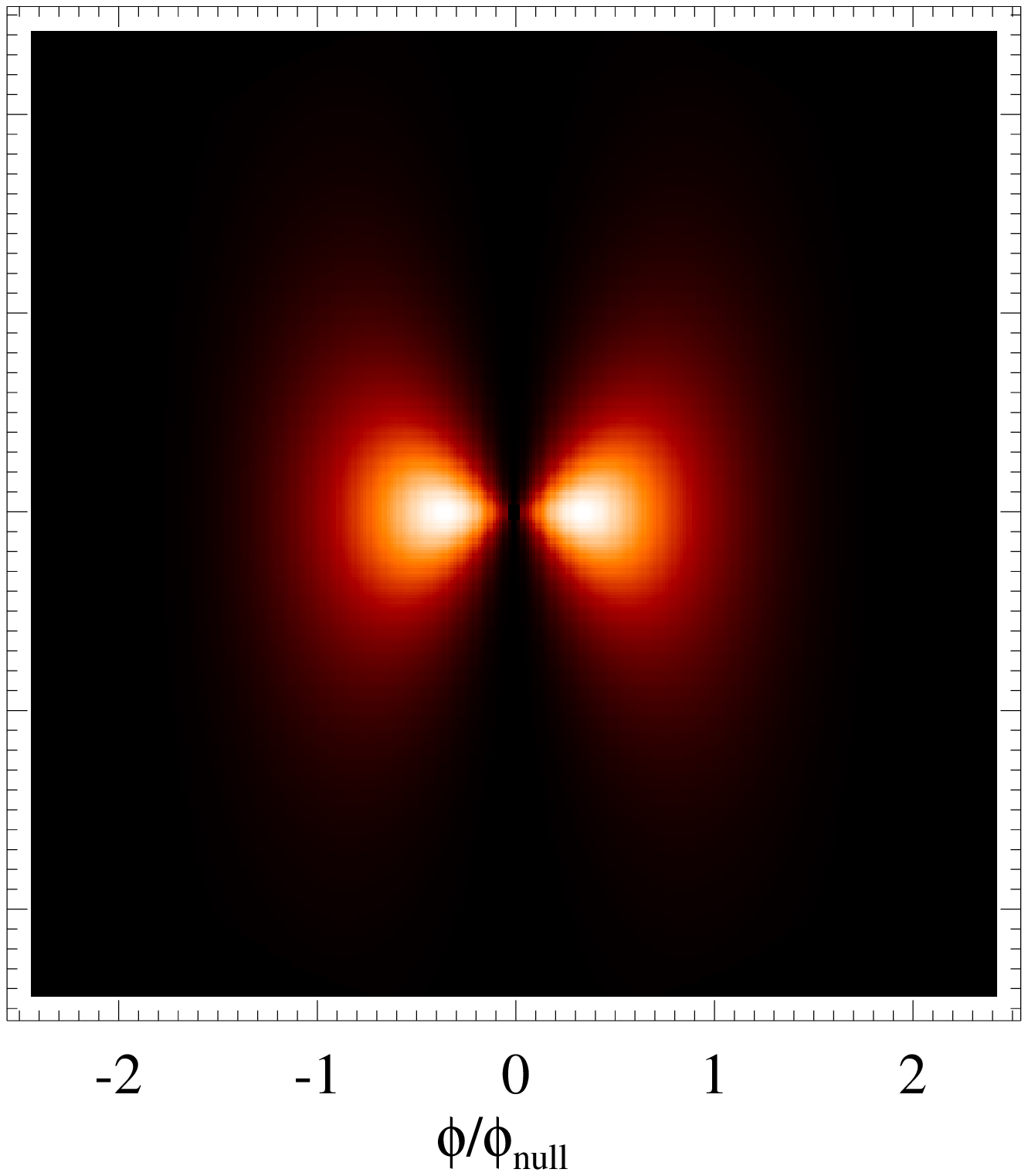}
    \caption{Illustration of LBTI transmission for a face-on disk observed at transit,
      where lighter regions correspond to greater disk surface brightness or greater LBTI
      transmission. All panels have the same spatial scale, but are labelled with
      different units (arcsec or $\phi_{\rm null}$). The left panel shows a face-on image
      of our disk model at 10pc (the star has been omitted). The middle panel shows the
      LBTI transmission pattern projected on the sky, with bright fringes being
      transmission maxima. The convention for the angle $\Omega_{\rm LBTI}$ is also
      shown. The right panel shows the transmitted disk flux (i.e. the left panel
      multiplied by the middle panel). For real observations the LBTI detector sees the
      right panel convolved with the diffraction-limited beam of a single LBT mirror
      (with a FWHM of about 280 mas). The computed null depth is the total flux density
      in the right panel divided by the stellar flux density.}\label{fig:trx}
  \end{center}
\end{figure*}

The original concept for nulling interferometry was put forward by
\citet{1978Natur.274..780B}, and for a reasonably non-technical description of the
instrument we refer the reader to Defr\`ere et al. (submitted). Basically speaking, the LBTI
combines the beams from the two LBT mirrors, with one beam having a half-wavelength phase
shift. Light in the viewing direction, and along lines perpendicular to the mirror
baseline vector, is therefore suppressed. Light from off axis sources at odd multiples of
the angular distance $\lambda/(2B)$ is transmitted ($B$ is the baseline
length). Similarly, light from sources at even multiples of this angular distance is
suppressed. The transmission (or ``fringe'') pattern is therefore a $\sin^2$ function
parallel to the baseline vector, and is constant perpendicular to this vector. A model
disk, the transmission pattern, and the transmitted disk image are illustrated in
Fig. \ref{fig:trx}. This illustration is for an object at transit; at other times the
fringe pattern is not aligned with North.

The key measurement from an LBTI observation is the ratio of the transmitted flux to the
total photometric flux. Because the transmission pattern is designed to suppress light
from the star, leaving light from a much fainter disk, this ratio is approximately the
ratio of the transmitted disk flux to the stellar flux. Deriving the transmitted disk
flux is not straightforward however, as some on-axis (i.e. stellar) flux is always
transmitted because the instrument is not perfect and the star has a finite angular
size. Removal of these effects is an integral part of the data analysis (see Defr\`ere et
al, submitted), so the quantity derived from a given observation is the ``calibrated null
depth'' or ``source null depth''.  In what follows we generally refer to this measurement
and the same quantity derived from our models as simply the ``null depth''. Because the
total flux is almost exactly the stellar flux and dominates over any disk emission, we
therefore compute the null depth for our models as the transmission for a starless model
divided by the stellar flux density.

\subsubsection{Disk transmission calculation}

Our disk model is radially extended, but to illustrate how surface brightness is
transmitted through the LBTI transmission pattern as a function of angular scale we first
consider a series of discrete annuli of angular radii $\phi=r/d$. We later combine these
annuli to model the transmission for our disk models.

\begin{figure}
  \begin{center}
    \hspace{-0.5cm} \includegraphics[width=0.5\textwidth]{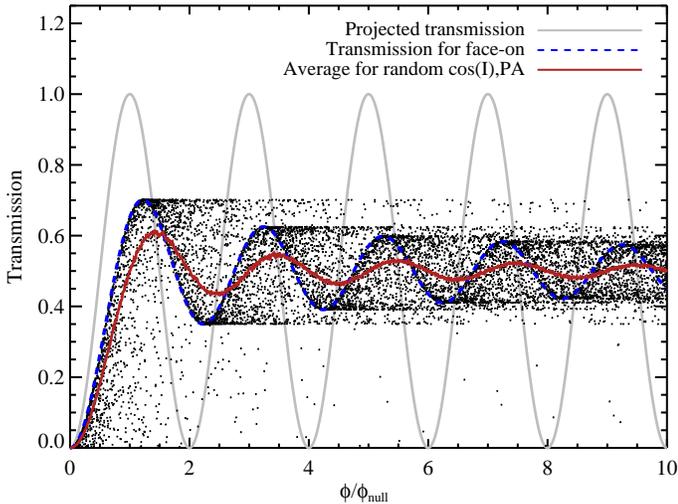}
    \caption{LBTI transmission for annuli of different angular sizes relative to the
      first transmission peak (where $\phi/\phi_{\rm null}=1$). The grey line is the
      projected transmission function parallel to the LBTI baseline, the blue dashed
      line is the fraction of emission for face-on ($i=0$) annuli after summing around
      the azimuthal angle. Dots show a population of annuli with random inclinations and
      position angles, and the red line shows the average transmission of these
      dots.}\label{fig:trvsr}
  \end{center}
\end{figure}

The angular distance to the first transmission peak is $\phi_{\rm null} = \lambda /(2B)$;
for the LBTI $\lambda=11$ $\mu$m and $B=14.4$m, so the distance is 79 mas. The
transmission function perpendicular to the fringes $T_{\rm null}$ is $\sin^2(\pi \phi/[2
\phi_{\rm null}])$, and is shown as the grey line in Figure \ref{fig:trvsr}, where the x
axis is the distance from the annulus center in units of the distance to the first
transmission peak (i.e. $\phi/\phi_{\rm null}$). A common definition for the inner
working angle of an instrument is where the sensitivity first reaches half of the peak
value, which for LBTI is therefore at $\lambda/(4B)=39$ mas.

To calculate the transmission for a point at some azimuth around the annulus requires
finding the sky-plane component of this vector point (relative to the star) that is
perpendicular to the fringe pattern. This component can be calculated using three
rotations, and is
\begin{equation}
  \phi_{\rm proj} = \phi \times \left( \sin \Omega_{\rm LBTI} \cos \theta + \cos \Omega_{\rm LBTI}
    \sin \theta \cos i \right)
\end{equation}
where $\theta$ is the angle from the sky plane around the annulus to the point of
interest, and $\Omega_{\rm LBTI}$ is the position angle of the annulus relative to the
LBTI fringe pattern. The angle $\Omega_{\rm LBTI}$ varies with hour angle, and because
the LBTI baseline is always perpendicular to the local vertical (i.e. a great circle
through the target and zenith), is equal to $\Omega$ for an object at transit. The
transmission from a point in the annulus is then\footnote{This transmission function is
  identical to the transmission at null given for the KIN by \citet{2011ApJ...734...67M}
  \begin{equation}
    T_{\rm null} = \left( 1- \cos \left[ 2 \pi \left( xu+yv\right) \right] \right)/2
  \end{equation}
  where $x,y$ are the sky offsets from the null center and $u,v$ the corresponding
  spatial frequencies (i.e. $xu+yv=\phi_{\rm y} B/\lambda$). LBTI transmission is simpler
  to compute because the length of the sky-projected baseline is always the same due to
  the common mount for the mirrors.}
\begin{equation}
  T_{\rm null} =  \sin^2( \pi \phi_{\rm proj} / [2 \phi_{\rm null}] ) \, .
\end{equation}
For a face-on ($i=0$) annulus at radius $\phi$, the transmission is clearly a function of
azimuth around the annulus. Averaging around an annulus yields the total transmission for
that annulus, and repeating this calculation for annuli of different angular sizes gives
the blue dashed line in Figure \ref{fig:trvsr}.
    
This exercise is finally repeated for a large number of annuli with random orientations,
so that $\cos i$ is distributed evenly from 0 to 1, and $\Omega_{\rm LBTI}$ is evenly
distributed from 0 to 180$^\circ$. Again, the transmission around each annulus is
azimuthally averaged, which results in the dots shown in Figure
\ref{fig:trvsr}. Averaging these points yields the average transmission as a function of
annulus radius for a population of disks with random orientations, shown as the red
line. As can be surmised from the decreasing amplitude, the average transmission tends to
0.5 at large separations (i.e. the average of $\sin^2$ is 0.5). The minimum separation at
which this transmission is achieved is $\phi_{\rm null}$, twice the inner working angle.

The origin of the dot distribution can be understood by considering how annuli of
different orientations are transmitted through a given transmission peak. For example,
the upper envelope of dots at about 70\% transmission are all transmitted through the
first transmission peak, and those at higher $\phi/\phi_{\rm null}$ are nearer to edge-on
with position angles closer to $\Omega_{\rm LBTI}=0^\circ$ (i.e. perpendicular to the
baseline and parallel to the fringes). This effect is relatively common because the
average inclination is about 60$^\circ$ (i.e. biased towards edge-on). By comparing the
phase of the grey line with the red line, it is clear that the peak average transmission
is actually about a quarter of the way beyond a transmission peak, and that the peak
transmission for face-on annuli lies somewhere in between. The phase shift of the face-on
transmission can be understood by realising that an annulus with a radius $\phi$ that is
slightly greater than $\phi_{\rm null}$ has more emission in the peak transmission region
than an annulus with $\phi=\phi_{\rm null}$. The average transmission for random
orientations is phase shifted slightly further because most disks are inclined, and
therefore on average appear somewhat smaller on the sky than they actually are.

\subsection{Transmitted disk flux}\label{ss:trx}

\begin{figure}
  \begin{center}
    \hspace{-0.5cm} \includegraphics[width=0.5\textwidth]{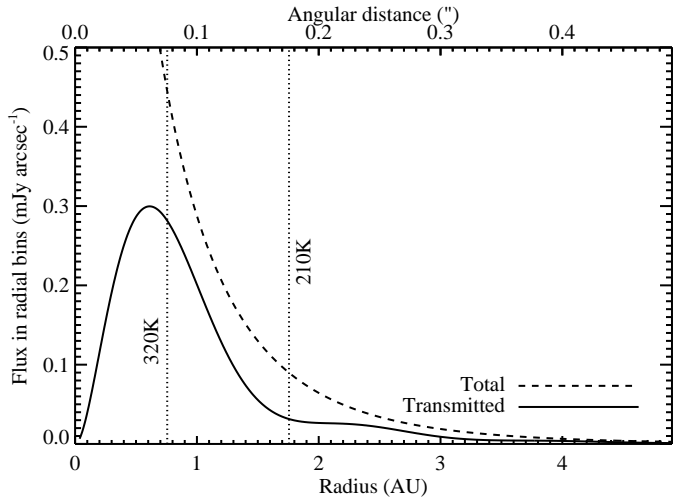}
    \caption{Total and transmitted flux density per unit radius, as a function of radius
      for a single face-on reference model disk at 10pc. This plot therefore shows
      azimuthally summed radial profiles from the left and right panels of
      Fig. \ref{fig:trx}. Flux is lost at all radii due to the transmission pattern, but
      most is lost from inside the habitable zone. The total and transmitted disk fluxes
      are 0.11 and 0.031 mJy.}\label{fig:fnutvsr}
  \end{center}
\end{figure}
 
Using the reference model of section \ref{sss:ref}, Fig. \ref{fig:fnutvsr} shows the
total and transmitted disk flux as a function of radius for a face-on geometry and a
distance of 10 pc. That is, the figure is a histogram showing where the total and
transmitted flux originates, so the solid line is an azimuthally summed radial profile
created from the right panel of Fig. \ref{fig:trx}. Much of the total disk emission
originates from the inner regions, and since the disk inner edge is well inside the first
transmission peak significant flux from these inner regions is not transmitted. Some flux
is also lost at larger radii, and in total only about 30\% of the disk flux is
transmitted. Overall therefore, the radial distance over which the disk emits strongly in
the mid-IR and can be detected is not particularly large; emission is strongly reduced
inside the inner working angle, and the faint Wien side of cooler emission means that the
surface brightness drops steeply at larger radii. Given the spacing of the transmission
peaks and the distance to nearby stars, the transmitted flux detected by the LBTI is
constrained to come from near the HZ, even if it cannot be certain that it originates
within it.

\begin{figure}
  \begin{center}
    \hspace{-0.5cm} \includegraphics[width=0.5\textwidth]{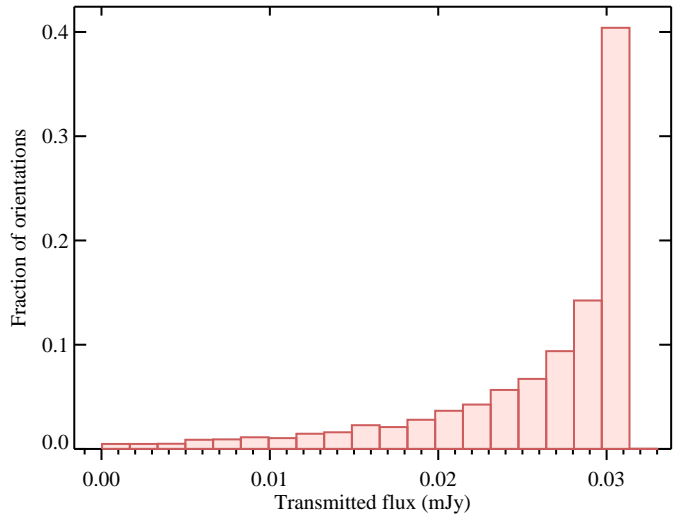}
    \caption{Distribution of transmitted fluxes for our reference model at 10 pc (with
      $z=1$) and randomly distributed orientations. The dotted line shows the average
      transmitted flux derived in Figure \ref{fig:fnutvsr}. The disk will sometimes
      appear fainter than average, but is often a little bit brighter. The mean level is
      0.026 mJy, and the medial level 0.029 mJy.}\label{fig:ftrhist}
  \end{center}
\end{figure}

Figure \ref{fig:ftrhist} shows how the total transmitted flux is distributed for a random
distribution of disk orientations. The concentration of transmitted fluxes near the
maximum arises because any disk with a position angle perpendicular to the transmission
pattern has the same transmitted flux, regardless of inclination (and any disk near to
face-on also has the same transmitted flux). The low transmitted flux tail arises from
disks that have a position angle parallel to the transmission pattern and are
sufficiently near to edge on such that almost the entire disk lies in the central null
transmission region. Such a geometry is relatively unlikely, as can be seen by the lack
of dots with low transmission in Fig. \ref{fig:trvsr}. These relatively rare low
transmitted fluxes are allowed because we assume disks with negligible scale height. This
distribution therefore represents a pessimistic (but possible) case, and becomes tighter
as the disk scale height increases. The width of the distribution also depends on the
range of hour angles over which a target is observed, as discussed in section
\ref{ss:ha}.

The point here is that the transmitted flux depends on the orientation, and since this is
expected to be unconstrained for LBTI targets (unless we use stellar inclination or the
orientation of a resolved outer disk as an estimate, which will be possible for some of
the HOSTS sample), there will be a corresponding uncertainty in any disk parameters that
are derived from the observations.

Since the transmitted flux scales linearly with $\Sigma_{\rm m,0}$, the distribution of
$\Sigma_{\rm m,0}$, and hence $z$, required to reproduce a given transmitted flux is
directly related to that in Figure \ref{fig:ftrhist}. The distribution of possible null
depths for our reference model at 10 pc is therefore calculated by dividing the
distribution of transmitted fluxes in Fig. \ref{fig:ftrhist} by the stellar flux. The
distribution of zodi levels $z$ implied for a given observed null depth (or upper limit)
is found by dividing the observed null depth by the distribution of model null depths for
$z=1$. This was the method employed by \citet{2011ApJ...734...67M} in modelling KIN
observations with \zp.

\subsection{Sky rotation}\label{ss:ha}

Any real LBTI observation takes a finite amount of time, during which any disk will
rotate relative to the transmission pattern on the sky. Thus, even a vertically thin
edge-on disk, which can be instantaneously invisible to the LBTI, will be visible when the
length of the observation is taken into account.

To compute how the disk transmission changes for an observation of a given object
requires computing the position angle of the LBTI fringe pattern on the sky at a given
hour angle. The common mirror mount for the LBT means that this angle is the difference
between the vector in the direction of the local vertical and a vector pointing towards
equatorial North along a line of constant right ascension through the object in question
(i.e. the hour circle). This angle is commonly called the parallactic angle.

\begin{figure}
  \begin{center}
    \hspace{-0.5cm} \includegraphics[width=0.5\textwidth]{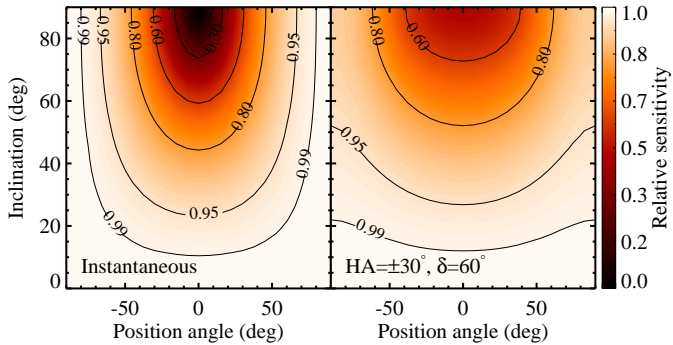}
    \caption{LBTI sensitivity relative to a face-on disk for instantaneous (left) and 4
      hour (right) observations (the changing sensitivity is purely due to the changing
      position angle of the disk, not the time needed to reach a given S/N). An edge-on
      disk with $\Omega=0^\circ$ is not instantaneously detected, but is only about 50\%
      fainter than a face-on disk for a 4 hour observation.}\label{fig:orient}
  \end{center}
\end{figure}

The null depth is then computed as the average null depth over the range of observed hour
angles. For modelling purposes here, these are spaced evenly in hour angle, but future
models will use the true distribution for each specific
observation. Fig. \ref{fig:orient} shows the difference between the instantaneous
sensitivity for a transiting target (left panel) and a 4 hour observation (right panel),
for a target at $60^\circ$ declination over the parameter space of different disk
orientations. This calculation is purely related to the changing position angle of the
disk, and does not represent the time needed to reach a given S/N. Here, as before, we
have assumed a vertically thin disk. The dark region in the upper middle of the left
panel shows that an edge on disk is not instantaneously detectable when the disk is
aligned with the LBTI fringe pattern. In the right panel, the rotation of the fringe
pattern with respect to North over a four hour observation means that the sensitivity
relative to a face-on disk is reduced by about 50\% for the worst case disk orientation.

\begin{figure}
  \begin{center}
    \hspace{-0.5cm} \includegraphics[width=0.5\textwidth]{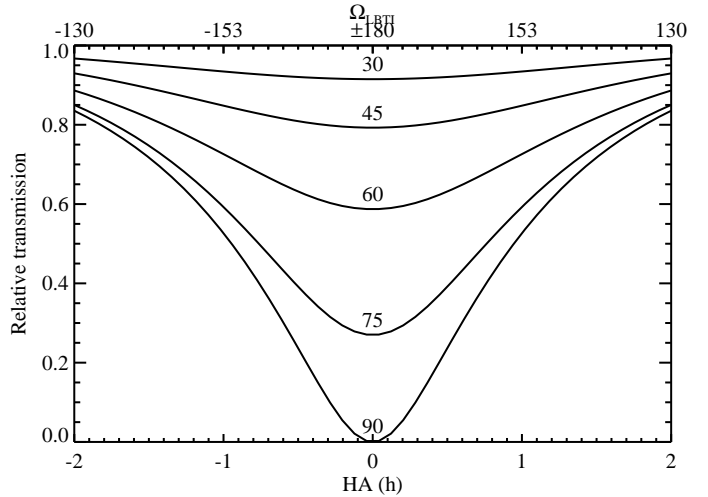}
    \caption{Instantaneous transmission through the LBTI over a range of hour angles for
      disks with $\Omega=0^\circ$, relative to a face-on disk. Each line shows the
      transmission for a different disk inclination. The top axis shows the disk position
      angle relative to the fringe pattern, which is 180$^\circ$ at transit because the
      LBT is facing North at that time. The curves at transit (HA=0) yield the
      transmission-inclination relation shown in the left panel of Fig. \ref{fig:orient}
      at $\Omega=0$, while the average value of each curve yields the
      transmission-inclination relation shown in the right panel. The curve average is
      always greater than the transmission at transit, so observations over a wider range
      of hour angles can detect disks with less favourable
      orientations.}\label{fig:orient_ha}
  \end{center}
\end{figure}

To understand the origin of this increased sensitivity, Fig. \ref{fig:orient_ha} again
shows the transmission relative to a face-on disk, but now for $\Omega=0$ at each point
in the range of hour angles from -2 to 2h. Each curve shows how the transmission for a
different disk inclination varies with hour angle (or $\Omega_{\rm LBTI}$). At HA=0, the
curves correspond to a cut at $\Omega=0$ in the left panel of Fig. \ref{fig:orient}. The
average of each line over the hour angle range corresponds to the same cut, but in the
right panel. Because the average of each curve is greater than the minimum value at HA=0,
the disk sensitivity is greater.

Of course, the inclination and position angle dependence in Figs. \ref{fig:orient} and
\ref{fig:orient_ha} shows that it would be highly desirable to discern how the
transmitted disk flux changes as a function of hour angle in order to learn about the
disk geometry and the possibility of non-axisymmetric structure. For axisymmetric disks
there is also a potential degeneracy between the disk geometry and the vertical scale
height; an edge-on disk with sufficient vertical extent will vary much less than the
curves in Fig. \ref{fig:orient_ha}, and may be indistinguishable from a less inclined
disk that is oriented such that the transmission variation is small. If a large variation
in transmission is seen, the disk may be vertically thin, but could alternatively be
non-axisymmetric. The detection of such effects will certainly be sought, but given that
we expect most LBTI detections to be near the sensitivity limits and therefore at low S/N
we assume for now that LBTI observations can be modelled by simply averaging over the
hour angle.

\subsection{Linking LBTI to other observations}\label{ss:other}

The LBTI is well suited to observing dust levels in the habitable regions around nearby
stars, both in terms of baseline length and observing wavelength. LBTI results will
therefore broadly address the goal of characterising warm dust levels without requiring
additional information from other observations. However, the specific location and radial
extent of the dust will be poorly constrained, and as illustrated by
Fig. \ref{fig:fnutvsr}, significant disk emission can arise from regions both inside and
outside the habitable zone and still be detected with the LBTI. In general, such
degeneracy in disk models will be hard to break because the goal of the LBTI is to detect
disks that are fainter than current detection limits (hence the need for the modelling
framework outlined in this paper).

In some cases however, LBTI results will be compared to other observations, most likely
(spectro)photometric detections or limits for the total disk flux density at the same
wavelength, and in some cases limits on disk size from high-resolution imaging or
interferometry. Detections and/or upper limits from near-IR interferometry will provide
constraints on dust levels inside the LBTI IWA.  LBTI will similarly provide very strong
constraints on the emission spectrum and location of hot dust detections
\citep{2013A&A...555A.104A,2014arXiv1409.6143E}.\footnote{Unless the dust has a very high
  albedo ($\gtrsim$0.9), non-detection with high quality mid-IR photometry is already
  sufficient to restrict this dust to lie well inside the habitable zone (Kennedy et al.,
  in prep).} Mid-IR photometric data will come from \emph{Spitzer} InfraRed Spectrograph
\citep[IRS,][]{2004ApJS..154....1W,2004ApJS..154...18H}, WISE observations
\citep{2010AJ....140.1868W}, or in the future the spectrometers on the Mid-Infrared
Instrument (MIRI) on the JWST, which cover the same wavelength range as the LBTI. Such
photometry will provide the most useful constraints on the disk location, so we now
briefly outline how the total and LBTI-transmitted disk fluxes vary with model
parameters, in particular the disk inner and outer radii.

\begin{figure}
  \begin{center}
    \hspace{-0.5cm} \includegraphics[width=0.5\textwidth]{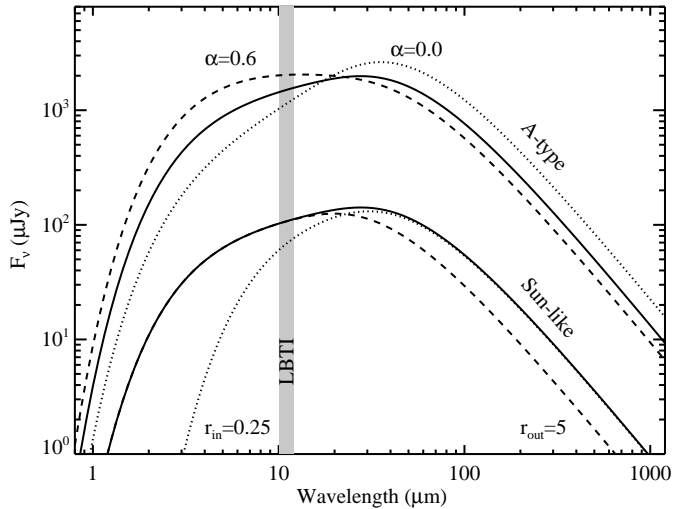} 
    \caption{Change in the disk spectrum with model parameters. The solid lines show the
      reference model for a Sun-like star (lower) and an A-type star (upper), both at 10
      pc. All disks have the same spectrum shape regardless of spectral type, but at
      fixed distance the total brightness increases to earlier spectral types that have
      physically larger disks. The dotted and dashed lines show how the spectrum changes
      for different values of $r_{\rm in}$ and $r_{\rm out}$, or for two different values
      of $\alpha$ (the defaults are $r_{\rm in}=0.034$ AU, $r_{\rm out}=10$ AU,
      $\alpha=0.34$). The LBTI N' band is shown as a grey stripe.}\label{fig:spec}
  \end{center}
\end{figure}

To illustrate the flux density distribution, Fig. \ref{fig:spec} shows disk spectra for
our reference model with $z=1$, with examples for both Sun-like and A-type host stars at
10 pc. Because our standard model has the same temperature range all disk spectra have
the same shape, but are brighter or fainter depending on the zodi level, the distance to
the host star, and the host star spectral type (which changes the area of the disk and
hence total brightness). The A5V star has $L_\star=14L_\odot$, so the disk around this
star is 14 times brighter than for the Sun-like star. However, the A-star is only about 5
times brighter at 11 $\mu$m (10 vs. 2 Jy), so the disk/star flux ratio has increased by a
factor of about 3.

Fig. \ref{fig:spec} also shows how the spectrum varies when $r_{\rm in}$, $r_{\rm out}$,
and $\alpha$ are changed from their standard values, and what the effect is at the LBTI
wavelength of 11 $\mu$m. Larger values of $\alpha$ concentrate the disk emission closer
to the star and hence shift the spectrum towards warmer temperatures. Changing $r_{\rm
  in}$ and $r_{\rm out}$ remove emission from the inner and outer disk regions, leading
to a spectrum that is closer to a single-temperature black body spectrum. As noted above
and shown by the curve for $r_{\rm out}=5$ AU, the outer disk radius is large enough that
the exact value matters little for the level of emission at 11 $\mu$m.

While our model gives a reasonable idea of the true disk spectrum, and could in principle
be used to constrain disk structure based on observations at other wavelengths, it is
inaccurate on a detailed level because grains do not emit like black bodies. Three
important differences are that i) the disk will be fainter than our model at long
wavelengths because grains emit inefficiently at wavelengths longer than their size, ii)
the spectrum may also be shifted to shorter wavelengths due to the presence of small
grains with temperatures greater than that of a black body, and iii) the spectrum may
show non-continuum spectral features. These differences pose problems for extrapolating
our model to other wavelengths, or at least mean that additional parameters such as grain
properties would need to be considered.

\begin{figure*}
  \begin{center}
    \hspace{-0.25cm} \includegraphics[width=0.5\textwidth]{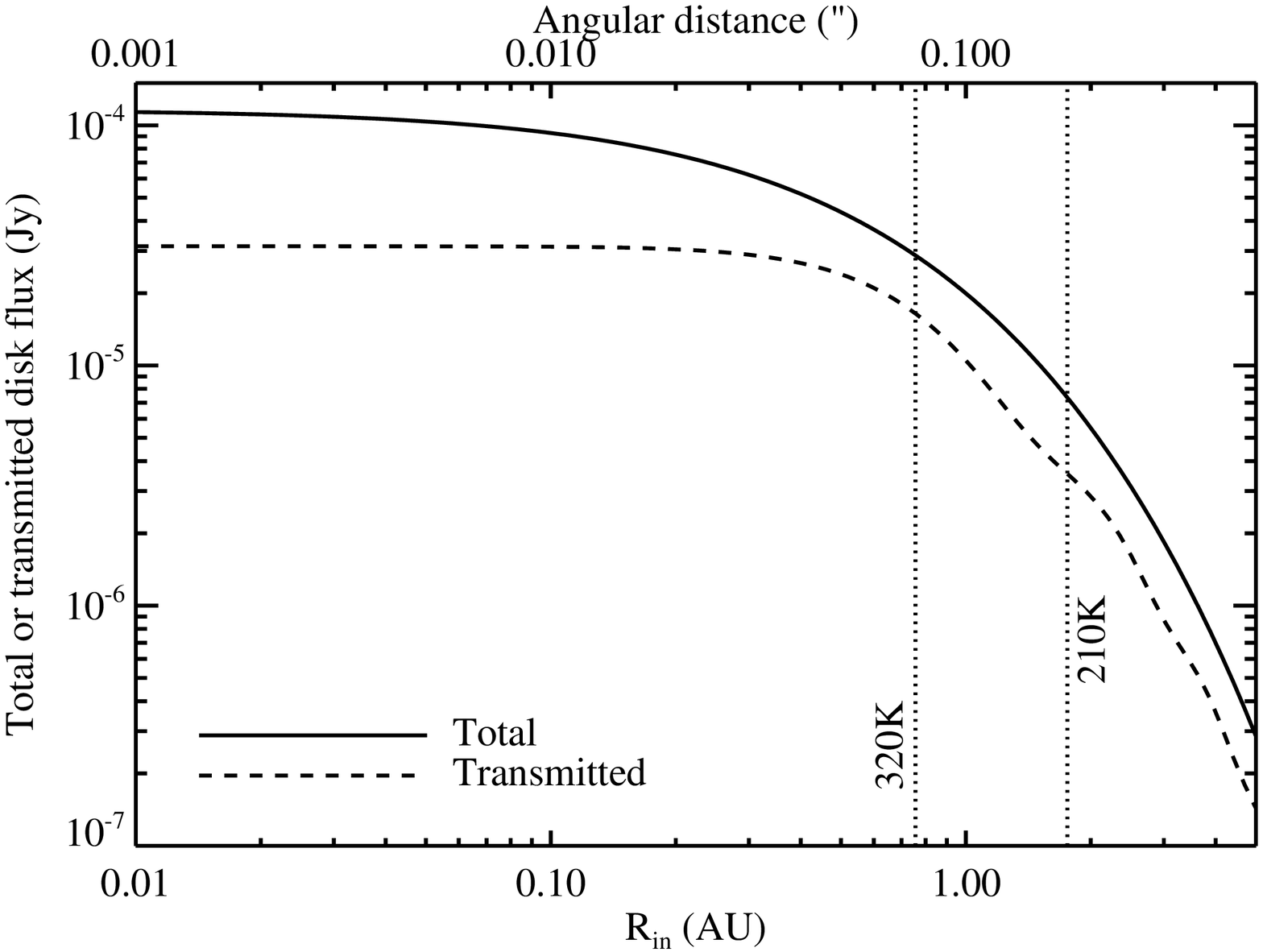} 
       \hspace{-0.75cm} \includegraphics[width=0.5\textwidth]{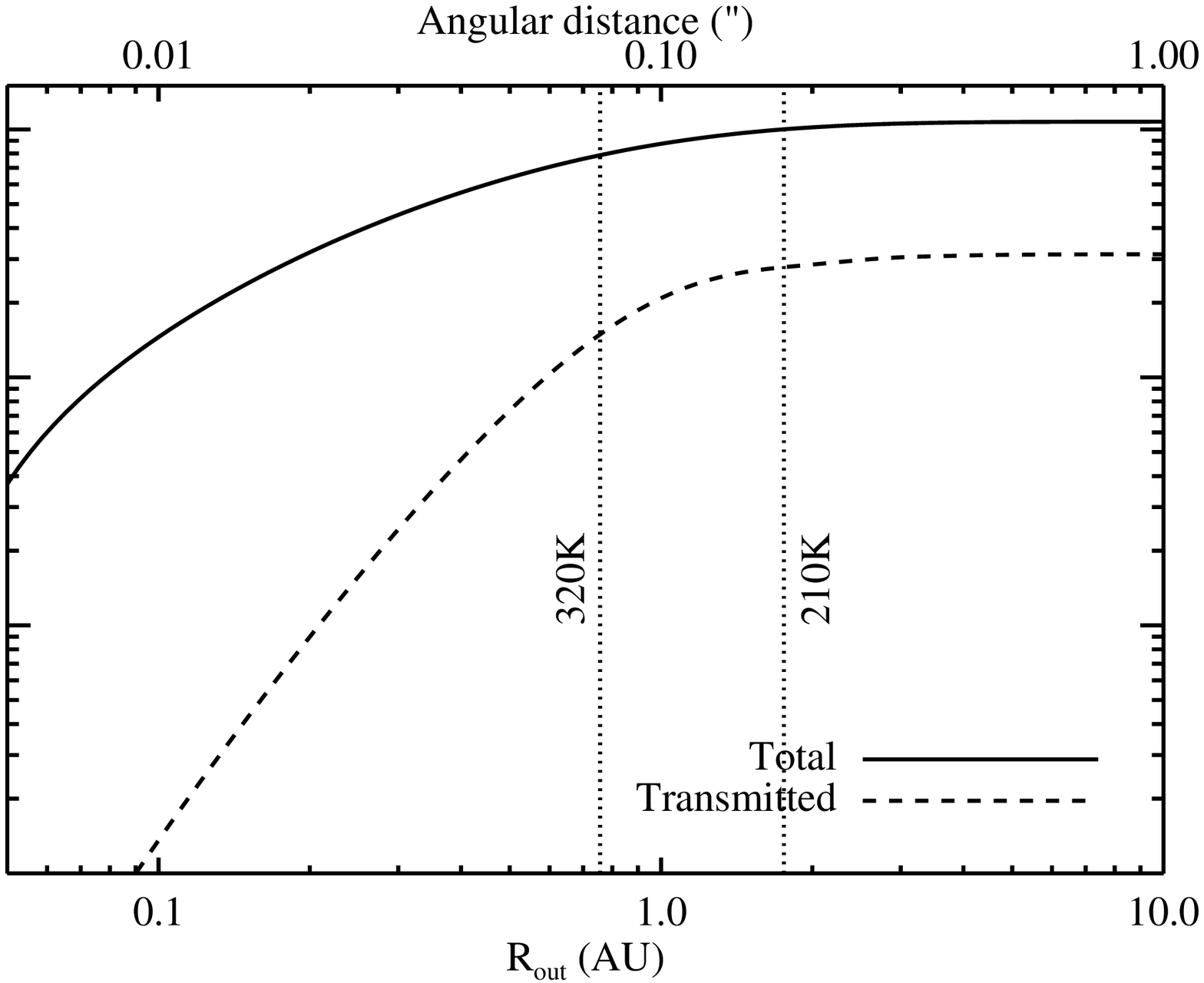} \\
       \caption{Change in total disk flux (solid lines) and transmitted disk flux (dashed
         lines) at 11$\mu$m as a function of $r_{\rm in}$ (left) and $r_{\rm out}$
         (right), for a face-on reference model disk around a Solar-type star at
         10pc. The total disk flux decreases as $r_{\rm in}$ increases, but is less
         sensitive to decreasing $r_{\rm out}$ because most of the mid-IR disk emission
         comes from the inner part of the disk (see also Fig. \ref{fig:spec}). The
         transmitted flux is insensitive to the disk extent when $r_{\rm in}$ lies inside
         the first transmission peak and $r_{\rm out}$ lies outside the first
         transmission peak (i.e. at about 0.7 AU at 10pc).}\label{fig:rin}
  \end{center}
\end{figure*}

Such differences are of minor importance here however, as important comparisons with
other observations can be made at the LBTI wavelength, so are largely independent of the
disk spectrum. The most important comparison is between the total and transmitted disk
fluxes, or equally the total disk to star flux ratio and the null depth. To illustrate
why combining such measurements is important, Fig. \ref{fig:rin} shows how the total and
transmitted model disk flux changes at 11 $\mu$m as the disk inner and outer radii
change. As $r_{\rm in}$ is increased the total disk flux decreases because hot emission
is being removed. However, the LBTI-transmitted flux changes little while $r_{\rm in}$
remains small, because these changes occur behind the central transmission minimum and
are invisible. When $r_{\rm in}$ is outside the first transmission peak (at about 0.7 AU
here) both the total and transmitted fluxes decrease in the same way. As $r_{\rm out}$ is
decreased there is initially little difference in the total and transmitted fluxes
because the outer disk is faint at 11 $\mu$m, but when $r_{\rm out}$ moves behind the
central transmission minimum the flux drops much more steeply than the total flux. As
noted earlier, Fig. \ref{fig:rin} shows that the LBTI transmission is insensitive to our
choice of disk inner and outer radii, as long as the disk is much wider than the
habitable zone.

Given such relations between the total and transmitted disk fluxes, it is clear that
observations of both may constrain the disk location. However, the transmitted disk flux
is a function of both the disk size and the orientation, so the best constraints on the
disk size require the orientation to be known (or take some assumed value). It is
unlikely that the position angle will be inferred from LBTI measurements in many cases,
so the disk orientation would probably be assumed based on other system information, such
as coplanarity with a resolved outer cool disk component, the known inclination and/or
position angle of the host star's rotation axis
\citep{2009A&A...498L..41L,2014MNRAS.438L..31G}, or of planet orbits
\citep{2009A&A...503..247R}, or a combination of all three
\citep{2013MNRAS.436..898K}. See Defr\`ere et al. (submitted) for an application of this
assumption to $\eta$ Crv.

\subsection{Scattered light surface brightness}\label{ss:scat}

Currently NASA is focussed on developing concepts for three optical (0.4-1 $\mu$m)
exoplanet-imaging missions, Exo-C (a coronagraph), Exo-S (a starshade), as well as an
ambitious precursor, the coronagraph on WFIRST-AFTA. Therefore, the thermal dust emission
information from the LBTI must be converted to predict the impact on scattered light
imaging. We now derive a simple prescription for converting the dust levels in the above
model into scattered light surface brightness estimates.

Scattered light predictions are in general difficult, and have largely proven
unsuccessful to date \citep[e.g.][]{2010AJ....140.1051K}, with debris disks imaged in
scattered light generally seen to be much fainter than predicted based on theoretical
grain models that match observed thermal emission. The typical minimum grain size in
debris disks is thought to 1-10 $\mu$m for Sun-like stars
\citep[e.g.][]{1994AREPS..22..553G,2010RAA....10..383K,2014ApJ...792...65P}, and the
steepness of the size distribution means that these grains dominate the surface
area. Such grains are expected to scatter optical light fairly isotropically, and have
fairly large albedoes of $\gtrsim$0.5. The scattered light faintness of debris disks,
where albedoes of 0.05-0.1 are seen
\citep[e.g.][]{2005Natur.435.1067K,2011AJ....142...30G,2007prpl.conf..573M}, likely arise
from incorrect assumptions about grain properties and sizes, and how these grains scatter
starlight.

The properties of exo-zodi are not sufficiently well known that considering physically
motivated possibilities for different grain sizes and properties would make predictions
for their scattered light brightness more certain. Therefore, regardless of the physical
reason, we will assume that the $\sim$0.1 effective albedoes seen for scattered light
disks around nearby stars are representative, and assume isotropic scattering. It may be
that the properties of warm dust are different to those inferred from scattered light
detections, since for example hot dust detected with near-IR interferometry is generally
inferred to originate in much smaller grains
\citep[e.g.][]{2011A&A...534A...5D,2013A&A...555A.146L}. For higher/lower albedoes, our
scattered light predictions would be higher/lower by the same factor.

While the model surface density $\Sigma_{\rm m}$ is connected to the true optical depth
and the surface area of grains from which the emission arises, the model surface density
is the true optical depth only if the grains behave like black bodies. That is, real
grains reflect starlight, so the true total surface area in the disk is always higher
than $\Sigma_{\rm m}$. To be more realistic, and to allow for the possibility of
scattered light, the grain absorption and scattering properties need to be considered.

Consider a disk with particles with a range of sizes $D$, and $\Sigma_{\rm true}(D) dD$
the true cross-sectional area per unit area of particles in the size range $D$ to
$D+dD$. Using the absorption efficiency as a function of size and wavelength $Q_{\rm
  abs}(\lambda,D)$, the thermal emission is
\begin{equation}
  S_{\rm th} = 2.35 \times 10^{-11} \int \Sigma_{\rm true}(D) Q_{\rm abs}(\lambda,D) B_\nu (\lambda,T[D]) dD.
\end{equation}
Though the dependence is not included explicitly here for simplicity, all quantities in
this equation also vary with location in the disk due for example to the disk structure
and changing composition. Using the scattering efficiency $Q_{\rm sca}(\lambda,D)$ the
scattered light emission can be written in various ways, but a convenient form is
\begin{equation}
  S_{\rm sca} =   \frac{F_{\nu,\star}}{4 \pi} \left(\frac{d}{r}\right)^2 \int \Sigma_{\rm true}(D) Q_{\rm sca}(\lambda,D) dD,
\end{equation}
which we could also express in terms of albedo $\omega = Q_{\rm sca}/(Q_{\rm abs} + Q_{\rm
  sca})$ by substituting $Q_{\rm sca} = Q_{\rm abs} \omega / ( 1-\omega)$.
If we assume the albedo $\omega$ is the empirical value of 0.1, independent of grain size
and wavelength, then because $Q_{\rm abs} = 1 - \omega$ the surface density $\Sigma_{\rm
  m}$ is approximately the true optical depth, but is underestimated by a factor of
$1-\omega$ (i.e. $\Sigma_{\rm m}=\Sigma_{\rm true} [1-\omega]$). That is, with these
assumptions the thermal surface brightness could be written
\begin{equation}
  S_{\rm th} = 2.35 \times 10^{-11} \Sigma_{\rm true} B_\nu(\lambda,T_{\rm BB}) (1-\omega) .
\end{equation}

The scattered light brightness is therefore calculated using the dust surface density and
the empirical effective albedo.
With these assumptions, the predicted scattered light emission from the model of section
\ref{ss:model} would be
\begin{equation}\label{eq:ssca}
  S_{\rm sca} = \frac{F_{\nu,\star}}{4 \pi} \left(\frac{d}{r}\right)^2
  \frac{\omega}{1-\omega} \Sigma_{\rm m} .
\end{equation}
For simplicity we use the value of $\Sigma_{\rm m}$ derived from our modelling, and thus
include an extra factor of $1/(1-\omega)$. Therefore, with our model the scattered light
surface brightness relative to the stellar flux\footnote{The dust surface brightness
  relative to the stellar flux is the relevant quantity here, because the light scattered
  from a putative exo-Earth will scale with the stellar flux in the same way as the
  dust.} in the habitable zone (i.e. $r=r_0$) of a 1 zodi disk decreases as $1/L_\star$,
as the habitable zone is pushed farther from the star by the increased luminosity. This
is the approach we use below to estimate the limiting scattered light surface
brightnesses for HOSTS, and for the LBTI $\eta$ Crv detection.

An extension to this approach would be to add a phase function that accounts for forward
scattering properly. The albedo could then be derived theoretically and calibrated with
observations. While in the above case the scattered light can be calculated simply along
the radial direction, and for a given orientation turned into an image, use of a phase
function $g(\theta)$ requires a three-dimensional calculation that includes the
star-particle-observer scattering angle $\theta$ at each location in the disk. Given that
we expect relatively large uncertainties in disk parameters, even in the case of a
detection, use of a detailed three-dimensional calculation including (assumed) grain
properties and a phase function will in general be unwarranted.

To convert the simple scattered light prediction of equation (\ref{eq:ssca}) to an
observable for an arbitrary disk inclination a three dimensional calculation must be made
to account for brighter disk ansae. However, the disk inclination will in general be
unknown, so a simple approximation for deriving a representative scattered light surface
brightness would be to assume an average inclination of $4/\pi$, and for a vertically
thin disk an increase of $1/\cos(4/\pi) \approx 3$.

\section{Modelling LBTI observations}\label{s:obs}

We now show two examples using the model described above. We first use the LBTI
commissioning measurement for $\eta$ Crv to illustrate how zodi levels are derived, and
then derive zodi limits for the HOSTS survey. These levels assume that the disk
orientation is not known, but in some cases coplanarity with a resolved outer disk could
be assumed to further constrain the disk brightness. This assumption can be made for
$\eta$ Crv, and we refer the reader to Defr\`ere et al. (submitted) for an in-depth
discussion of these results for this system.

\subsection{Generic zodi level calculation}


\begin{figure*}
  \begin{center}
    \hspace{-0.5cm} \includegraphics[width=0.384\textwidth]{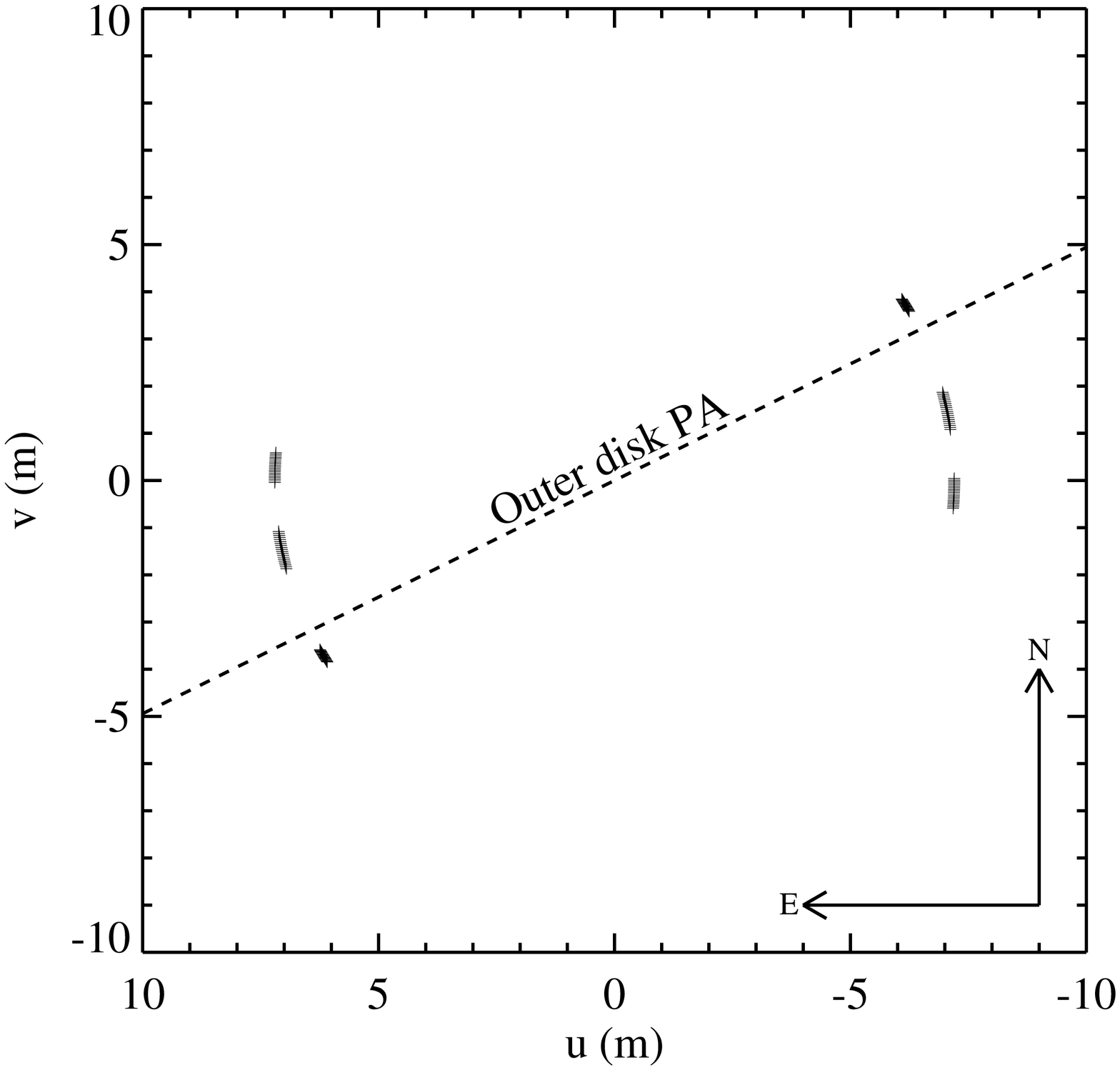}
    \hspace{0.5cm} \includegraphics[width=0.5\textwidth]{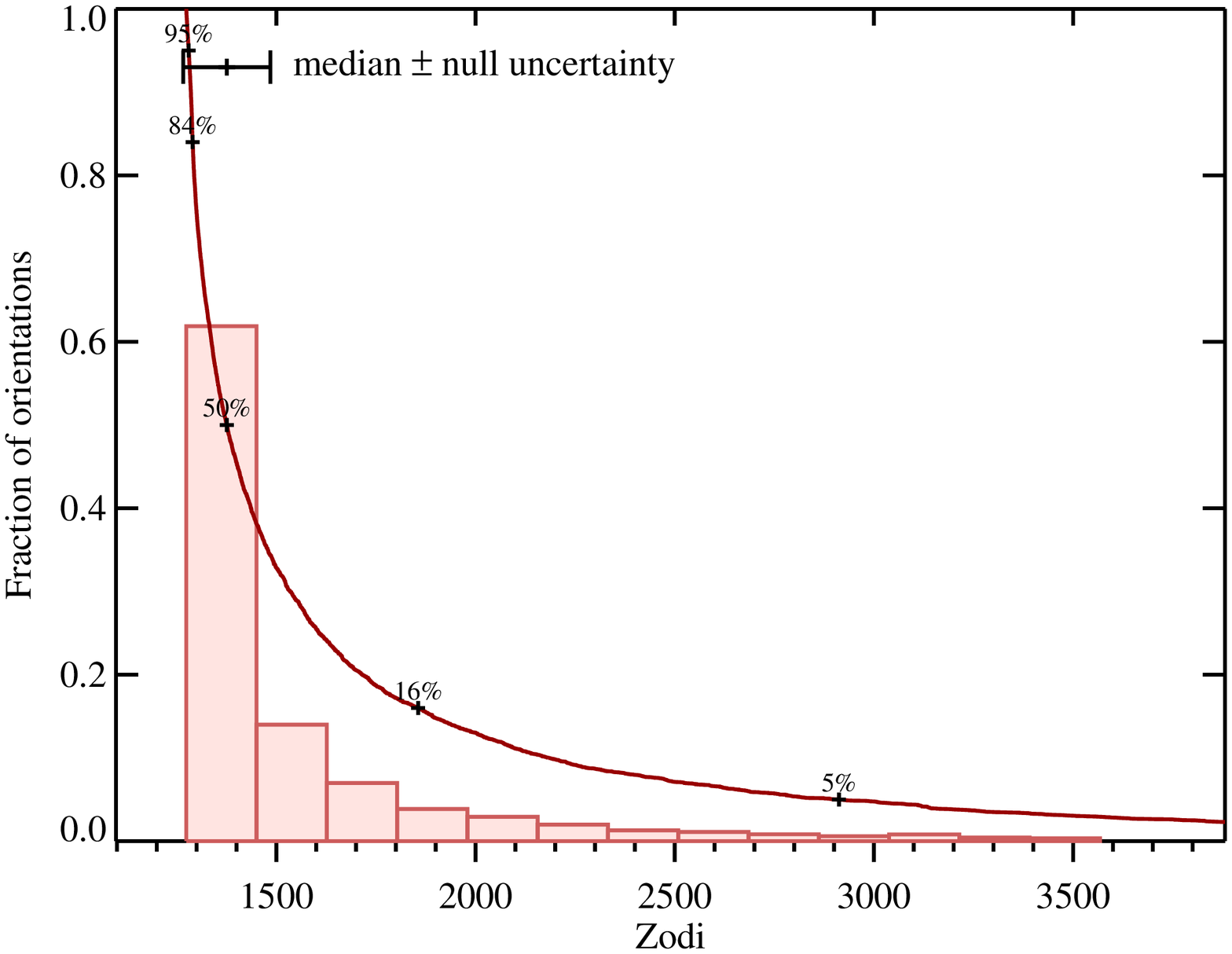}
    \caption{LBTI observation and zodi distribution for $\eta$ Crv. The left panel shows
      the uv coverage and outer disk position angle. The right panel shows the
      distribution of zodi levels for 5000 random disk orientations. The cumulative
      distribution of zodi levels is shown by the grey line. The median zodi level and
      the uncertainty purely due to the null depth uncertainty is also
      shown.}\label{fig:etacrvlim}
  \end{center}
\end{figure*}

Using $\eta$ Crv with a null depth of $4.4 \pm 0.35$\% as an example, we first show how
the distribution of zodi levels is derived assuming an unknown disk orientation. This
procedure is generic in that it can be applied to any LBTI observation, whether a
significant detection or an upper limit was found. The LBTI is still in commissioning so
the zodi sensitivity derived here is not illustrative of the expected performance.

Fig. \ref{fig:etacrvlim} shows the $uv$-plane coverage of the LBTI observation relative
to the disk position angle, and the distribution of zodi levels calculated using our
reference model. We include the rotation of the disk relative to the fringe pattern
during the observation to compute the average null depth for each disk orientation. As
described at the end of section \ref{ss:trx} the distribution is essentially the inverse
of that shown in Fig. \ref{fig:ftrhist}.

The distribution is strongly peaked at about $z=1350$, with a tail of larger values due
to unfavourable disk orientations. The 1$\sigma$ zodi uncertainty due purely to the
calibrated null depth measurement is given by the width of the ``median $\pm$ null
uncertainty'' error bar in the right panel of Fig. \ref{fig:etacrvlim}. Similarly, the 16
and 84\% levels from the cumulative distribution give a representative 1$\sigma$ range
due to the orientation distribution. The upper uncertainty on the zodi level is therefore
set by the orientation distribution, while the lower uncertainty is set by the null depth
measurement. The median zodi level for $\eta$ Crv is $z = 1376 \pm 102$ if the
uncertainty is set by the LBTI null depth measurement. Including the 1$\sigma$ range from
the orientation distribution in quadrature, the range covered on either side of the
median value is $1236 < z < 1869$. In general, we expect that the null depth uncertainty
will dominate the lower bound on the zodi level, and the orientation distribution will
dominate the upper bound.

For comparison, based on a KIN detection and using the \zp~model Mennesson et al. (in
press) found $z = 1813 \pm 209$ for $\eta$ Crv at 8.5 $\mu$m with no assumptions about
the disk orientation. Our derived zodi level is different for two reasons: i) $\eta$ Crv
is hotter than the Sun (6900K), so our luminosity-dependent zodi definition will lead to
a zodi level about 1.7 times smaller than \zp, and ii) the mid-IR spectrum of the $\eta$
Crv disk increases more steeply with wavelength than a black body (i.e. has a silicate
spectral feature), so although our model has nearly the same temperature profile as \zp,
our derived zodi level will be slightly larger because LBTI observes at 11
$\mu$m. Therefore, direct comparisons between our zodi levels and those using \zp~should
not be made.

\subsubsection{Generic scattered light surface brightness}

Using the simple prescription for the scattered light surface brightness described in
section \ref{ss:scat}, we can convert the zodi levels shown in Fig. \ref{fig:etacrvlim}
into a prediction for the face-on scattered light surface brightness. Here, the value of
interest is the surface brightness in the habitable zone, as this is where planets would
be sought. The level is then calculated using equation (\ref{eq:ssca}), with $\Sigma_{\rm
  m}=z \Sigma_{\rm m,0}$ because $\Sigma_{\rm m,0}$ is the value at the radial distance
where the equilibrium temperature is the same as Earth's. Adopting $\omega=0.1$, and
using $F_{\nu,\star}=72.35$ Jy in V band, $d=18.3$ pc, $r=\sqrt{L_\star/L_\odot}=2.3$ AU,
and $z=1372^{+497}_{-140}$ yields a scattered light surface brightness of $S_{\rm sca} =
4.0^{+1.4}_{-0.4}$ mJy/arcsec$^2$, or about 15 mag/arcsec$^2$. As noted above, this
habitable zone estimate may be increased by a factor of a few to account for the
inclination of the disk.


\subsection{Exo-Zodi detection limits for HOSTS}\label{ss:hosts}

Given a prediction for the sensitivity of the LBTI and a sample of stars that will be
observed, we can make predictions for the sensitivity of the HOSTS survey. In what
follows we assume a 1$\sigma$ uncertainty on the LBTI calibrated null depth of $10^{-4}$,
and hence the limits presented are also at 1$\sigma$. Limits are calculated as above,
using the median of the distribution of zodi levels over the random distribution of
orientations. Though we do not account for it here, there is some uncertainty in the
stellar flux densities and luminosities used, which we estimate to contribute at about
the 5\% level. Here we calculate the sensitivity assuming an observation at a single hour
angle, but when the model is applied to real observations in the future the calculations
will account for sky rotation as described in section \ref{ss:ha}.

These calculations are carried out for both our reference disk model, and a
``worst-case'' scenario where the dust is restricted to lie only within the habitable
zone, which we assume to lie between temperatures of 320 and 210K. As shown in
Figs. \ref{fig:fnutvsr} and \ref{fig:rin}, the LBTI is sensitive to emission that lies
interior and exterior to the habitable zone, so the dust surface density must be higher
in the latter case for an LBTI detection at the same sensitivity. As noted above in
section \ref{ss:zodi}, the ``zodi levels'' derived with this different radial structure
are also different and should really be considered as enhancements over the surface
density derived for the Solar zodiacal cloud, rather than zodi levels to be compared with
other values.

The HOSTS survey sample is described by Weinberger et al. (submitted), the key aspect
being that targets are chosen such that their habitable zones have larger angular sizes
than the first LBTI transmission peak, so observations directly probe the levels of
habitable zone dust. The sample is split by $B-V$ colour at 0.42 into ``Sun-like'' and
``sensitivity'' sub-samples, which simply reflects the levels of dust that can be
detected.

\begin{figure}
  \begin{center}
    \hspace{-0.5cm} \includegraphics[width=0.5\textwidth]{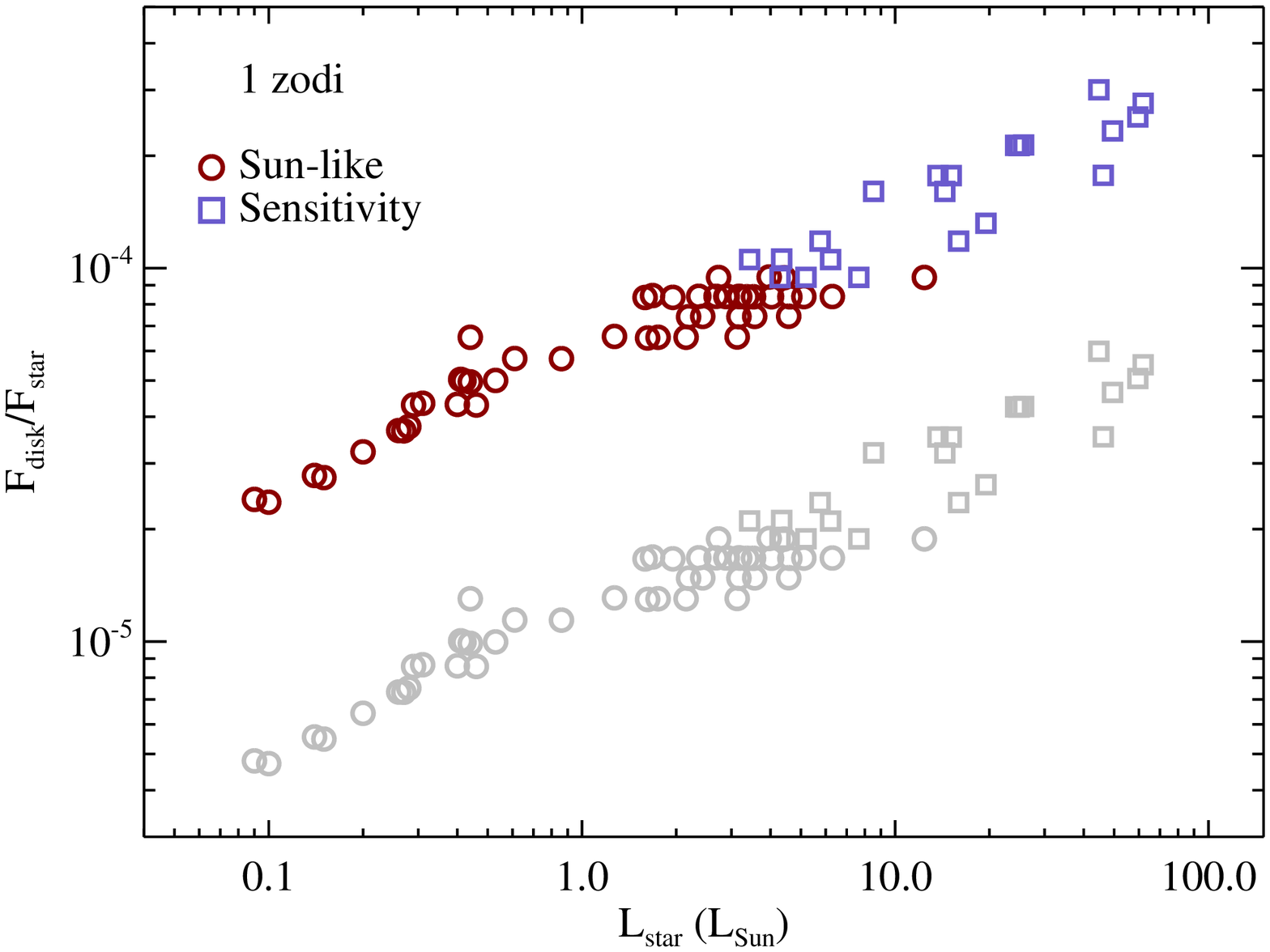} \\
    \hspace{-0.5cm} \includegraphics[width=0.5\textwidth]{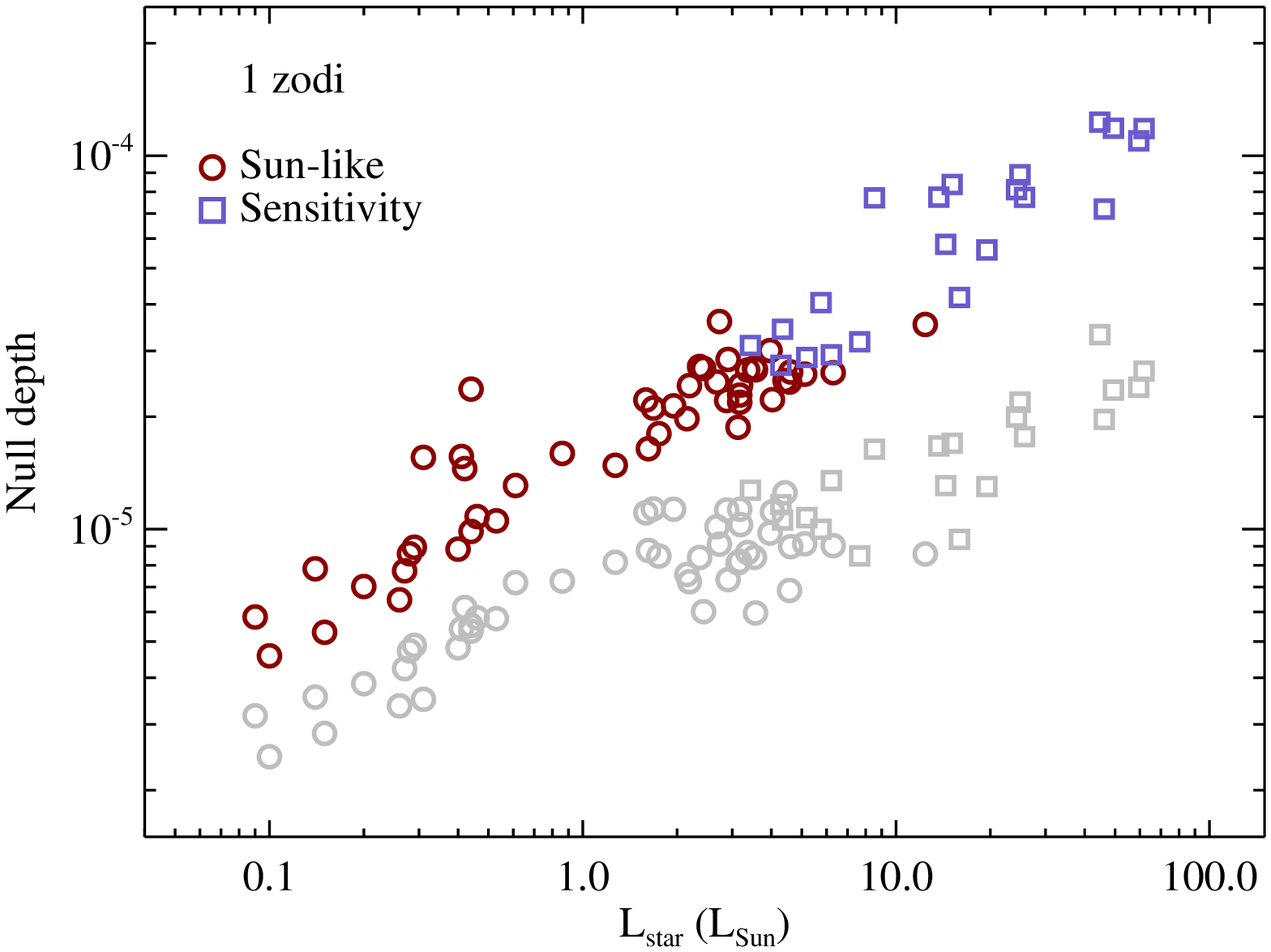} \\
    \hspace{-0.5cm} \includegraphics[width=0.5\textwidth]{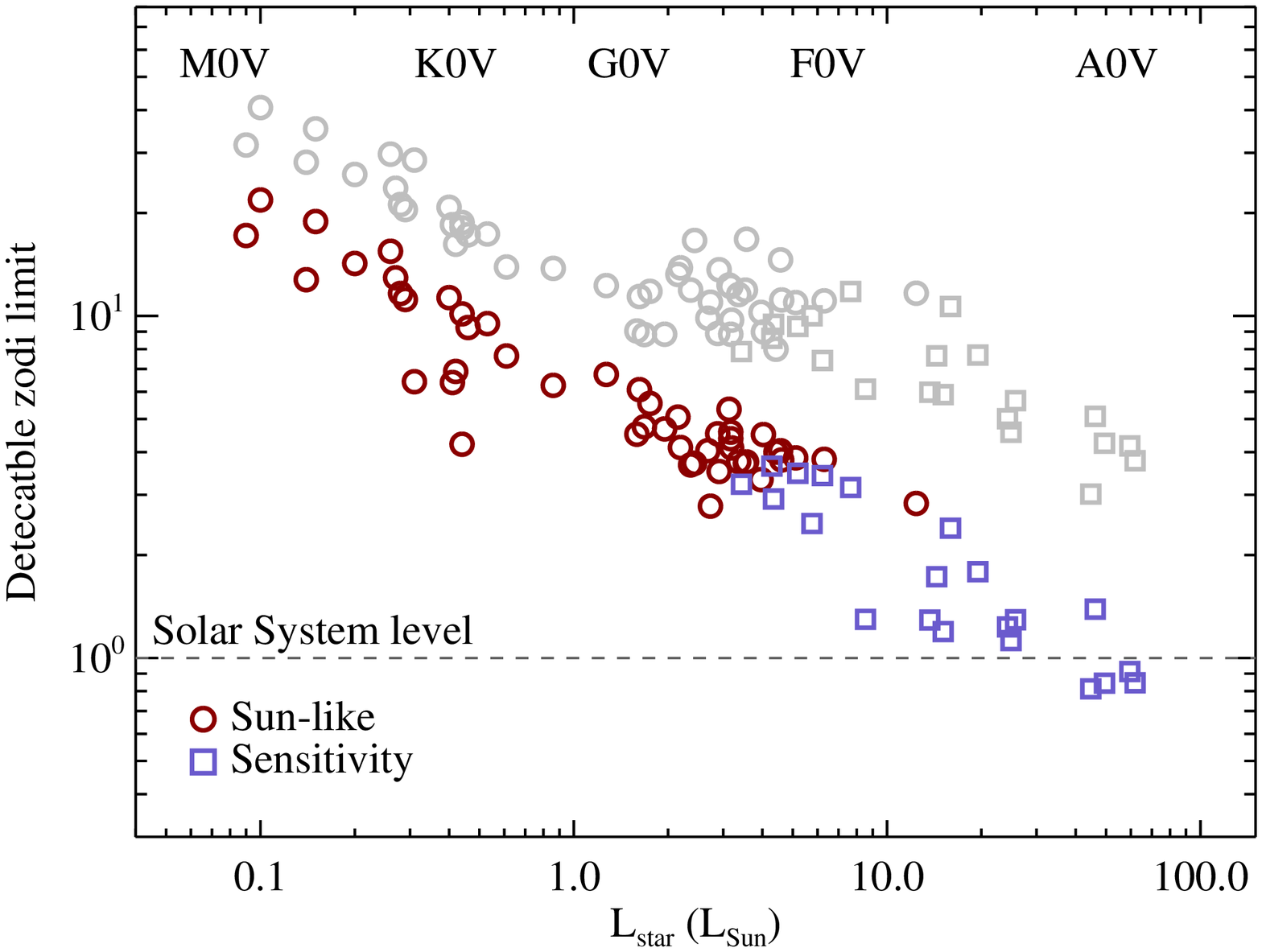}
    \caption{Disk to star flux ratio, null depth, and zodi limits for the reference disk
      model for HOSTS survey stars, split into ``Sun-like'' and ``sensitivity''
      sub-samples (red circles and blue squares). The top two panels show the flux ratio
      and null depth for a 1 zodi disk, and the bottom panel shows the sensitivity in
      zodis for the predicted LBTI sensitivity. The dashed line in the bottom panel shows
      the Solar System level. Grey symbols show a pessimistic narrow-disk scenario where
      disks only cover the habitable zone (from 320 to 210K), and the different disk
      width means that these values are not strictly zodi levels because they do not use
      our reference model (see section \ref{ss:zodi}).}\label{fig:leak}
  \end{center}
\end{figure}

Figure \ref{fig:leak} shows the predicted disk to star flux ratios, null depths, and
sensitivity in zodi units for the HOSTS sample using our reference model (red and blue
symbols) and our worst case scenario (grey symbols, which we discuss below). The top and
middle panels show the flux ratios and null depths expected for a 1 zodi model around
these stars, and the bottom panel shows the sensitivity in zodis for the predicted LBTI
sensitivity. There are clear trends with stellar luminosity, which can be understood as
follows. With our model the disk flux density at fixed wavelength scales with the angular
area, i.e., $\propto z \Sigma_{\rm m,0} L_\star/d^2$ (with our zodi definition
$\Sigma_{\rm m,0}$ is constant, but we include it to consider other zodi definitions
below).  The stellar flux, on the other hand, scales $\propto L_\star/(T_\star^3 d^2)$ in
the Rayleigh-Jeans regime so the total disk to star flux ratio only depends on the star,
and is higher for earlier spectral types
\begin{equation}\label{eq:fdfsmod}
  \frac{F_{\rm disk}}{F_{\nu,\star}} \propto z \Sigma_{\rm m,0} T_\star^3 \, .
\end{equation}
This dependence in the top panel of Fig. \ref{fig:leak} arises due to the stronger
scaling of stellar luminosity with temperature than flux density at 11 $\mu$m, with the
variations from a perfect correlation arising due to variation in $L_\star$ at fixed
$T_\star$ (i.e. different stellar radii). At fixed distance, as the stellar temperature
and luminosity increase the habitable zone is pushed outwards ($\propto \sqrt{L_\star}$)
and its area and brightness increase more rapidly than the stellar flux. Alternatively,
for a fixed disk angular size (and hence fixed disk brightness), increasing the stellar
temperature and luminosity pushes the system to greater distances and the star becomes
fainter due to increasing distance faster than it becomes brighter due to an increased
temperature.

The null depths for a 1 zodi disk are shown in the middle panel, and the trend has a similar
origin as the disk to star flux ratio. At fixed distance, increasing stellar luminosity
increases the disk surface brightness at fixed angular radius because $r_0$ increases
(i.e. increases if $\alpha$ is positive). Assuming that most of the transmitted disk flux
originates from a constant angular scale (i.e. near the first transmission peak), the
transmitted disk flux therefore increases as $\propto (\sqrt{L_\star}/d)^\alpha$
(assuming that the effect of the changing disk temperature in the first transmission peak is
small). Combining this expression with the stellar flux yields a null depth
\begin{equation}\label{eq:leakmod}
  {\rm null} \propto z \Sigma_{\rm m,0} T_\star^3 \left[ \left(\sqrt{L_\star}/d\right)^\alpha \right] \, .
\end{equation}
HOSTS stars are chosen to have habitable zones with similar angular sizes and for our
reference model $\alpha=0.34$, so the term in square parentheses varies relatively little
and the null depth almost entirely depends on the stellar temperature.

The absolute null depth level of course also varies linearly with the zodi level, so the zodi
limits can be derived by dividing the expected sensitivity of $10^{-4}$ by the null depth
values in the middle panel (i.e. by solving equation (\ref{eq:leakmod}) for $z$)
\begin{equation}\label{eq:zmod}
  z \propto \Sigma_{\rm m,0}^{-1} T_\star^{-3} \left[ \left(d/\sqrt{L_\star}\right)^\alpha \right] \, .
\end{equation}
The resulting sensitivities show that what we define as Solar System levels of zodiacal
dust are at the predicted 1$\sigma$ noise level for early-type stars, as are 3-10 zodi
disks around Sun-like stars.

This discussion of scalings applies equally to the narrow worst case scenario, shown as
grey symbols in Fig. \ref{fig:leak}. As expected for a narrower disk that emits over a
smaller total physical area, the disk to star flux ratios are lower than for our
reference model, as are the null depths. While the difference between the two models is
about a factor of five in disk/star flux ratio, the difference in null depths is only a
factor of two to three because most of the flux removed from the disk in the narrower
model is hidden behind the central transmission minimum
(e.g. Fig. \ref{fig:rin}). Similarly, the difference in zodi levels is a factor two to
three higher in this case compared to our reference model. Therefore, the effect of this
pessimistic scenario in terms of habitable zone dust levels is relatively minor.

Because our zodi definition is based on constant surface density in the habitable zone,
the $z$ dependence on luminosity directly shows that the LBTI can truly detect lower
surface densities of dust in the habitable zones of earlier type stars. This conclusion
does not depend on our zodi definition because the LBTI is sensitive to $\Sigma_{\rm
  m}(r_0)$. A different definition, for example $z \propto F_{\rm disk}/F_\star$, makes
the zodi limit approximately constant ($\propto [d/\sqrt{L_\star}]^\alpha$), but also
implies $\Sigma_{\rm m,0} \propto T_\star^{-3}$ (and the sensitivity to habitable zone
surface density is the same as with our definition).

\subsubsection{Scattered light for HOSTS}

\begin{figure}
  \begin{center}
    \hspace{-0.5cm} \includegraphics[width=0.5\textwidth]{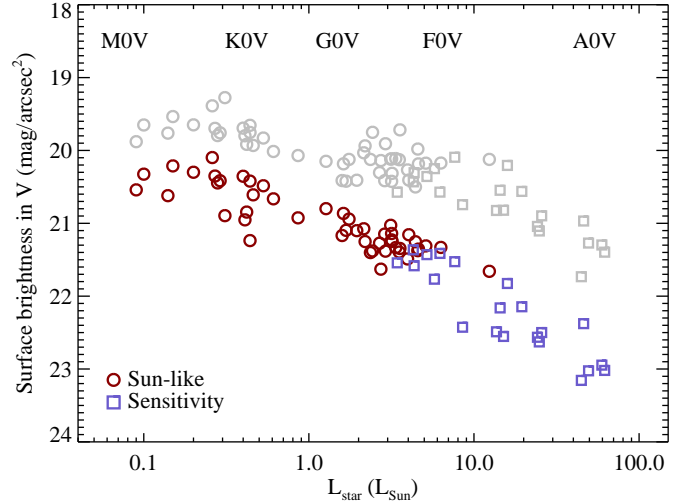}
    \caption{Face-on V band habitable zone scattered light surface brightness limits
      (assuming $\omega=0.1$) for the reference disk model for HOSTS survey stars, split
      into ``Sun-like'' and ``sensitivity'' sub-samples (red circles and blue
      squares). These limits use equation (\ref{eq:ssca}) and the zodi limits from
      Fig. \ref{fig:leak}. Grey symbols show a pessimistic narrow-disk scenario where
      disks only cover the habitable zone (from 320 to 210K).}\label{fig:ssca}
  \end{center}
\end{figure}

Given the detection limits in the bottom panel of Fig. \ref{fig:leak}, we can also derive
the face-on habitable zone ($r=r_0=\sqrt{L_\star/L_\odot}$ AU) scattered light surface
brightness of disks at the HOSTS survey detection limits. These limits are shown in
Fig. \ref{fig:ssca} for $\omega=0.1$ as defined in section \ref{ss:scat}. Red and blue
symbols show limits for our standard disk model, which again depend on the stellar
luminosity.

The origin of the dependence on stellar luminosity can be understood by rewriting
equation (\ref{eq:ssca}) using the zodi limit scaling above
\begin{equation}\label{eq:scamod}
  S_{\rm sca} \propto \frac{F_{\nu,\star}}{T_\star^3}
  \left( \frac{d}{\sqrt{L_\star}}\right)^{2+\alpha} \, .
\end{equation}
Like the zodi limits, the scattered light levels in the habitable zone at these limits
decrease with stellar luminosity. The squared distance and luminosity dependence is
simply the geometric effect that accounts for $1/r^2$ dilution of light and that surface
brightness is an angular measure. The extra $\alpha$ dependence is because the habitable
zone is not at the first transmission peak for all targets. If the LBTI observation is
dominated by emission near the first transmission peak, but the habitable zone is
slightly farther out (i.e. $d/{\rm 1pc}/\sqrt{L_\star/L_\odot} > 1$), the $\alpha$
dependence represents a model-dependent extrapolation that is relatively unimportant
while $\alpha$ is small.

As with the LBTI sensitivity to habitable zone surface density, our zodi definition has
no effect on the scattered light predictions and is merely an intermediate step for
deriving the true quantity of interest. That is, as described above the LBTI sensitivity
to $\Sigma_{\rm m}$ in the habitable zone (i.e. $z \Sigma_{\rm m,0}$), which sets the
limits on $S_{\rm sca}$, will always be the same. For the example of a constant $F_{\rm
  disk}/F_\star$ zodi definition therefore, $z$ is constant but $\Sigma_{\rm m,0} \propto
T_\star^{-3}$ and equation (\ref{eq:scamod}) is the same.

Because the HOSTS sample is explicitly chosen such that the LBTI is sensitive to thermal
emission from dust in the region of interest for Earth-imaging, the scattered light
predictions do not strongly depend on the choice of our disk model parameters (but of
course depends on albedo). The predictions are not totally independent of our disk model
however, since for example changing the steepness of the radial profile makes the dust in
the model more or less concentrated relative to the LBTI transmission pattern, thus
changing the derived zodi level. Varying $\alpha$ between -1 and 1 yields changes of
$\lesssim$1 mag in the scattered light predictions, meaning that the model dependent
uncertainty is similar to the variation expected from the unknown disk inclination.

For Solar-type stars, the scattered light levels in Fig. \ref{fig:ssca} for our face-on
reference model at the $\sim$4 zodi detection limit is 20-21 mag arcsec$^{-2}$. For
comparison, using a 4 zodi disk the \zp~model predicts 21.2 mag arcsec$^{-2}$ at 1 AU
with default parameters (a Solar analogue with $\omega=0.18$ and use of a phase
function), and 21.6 mag arcsec$^{-2}$ for $\omega=0.1$ and isotropic scattering. Our
simple model therefore compares well with the more complex calculation made by \zp. For a
60$^\circ$ inclined disk, \zp~gives 21.1 mag arcsec$^{-2}$, and for 90$^\circ$ gives 20.6
mag arcsec$^{-2}$, and therefore the increase from face-on to edge-on is about 1
magnitude. Our model has smaller vertical extent than \zp, so the increase in surface
brightness for inclined disks will be larger, roughly a factor of three for a 60$^\circ$
inclined disk.

The grey symbols in Fig. \ref{fig:ssca} show the scattered light surface brightness for
narrower disks that only lie in the habitable zone. These limits are about 1 magnitude
brighter than our reference model, so similar to the variation in brightness with the
(unknown) disk orientation. Therefore, the scattered light limits from LBTI observations
are fairly robust to disk width.

\section{Summary and Conclusions}

We have outlined a parameterised disk model to be used for modelling and interpreting
mid-IR exo-zodi observations with the LBTI. Using this model, we have illustrated how to
derive dust limits and levels for exo-zodiacal clouds, and how these can be converted to
scattered light surface brightnesses needed for planning future missions that will image
extrasolar Earth-analogues.

Using the HOSTS sample that the LBTI will observe, we illustrate the survey detection limits
both in terms of zodi units and the expected scattered light levels at these
limits. These limits are around ten times the Solar System level for Solar-type stars,
and thus the LBTI is expected to provide stringent limits with key information that will help
plan future Earth-imaging efforts.

\section*{Acknowledgments}

The Large Binocular Telescope Interferometer is funded by the National Aeronautics and
Space Administration as part of its Exoplanet Exploration Program. This work was
supported by the European Union through ERC grant number 279973 (GMK, OP, ABS \& MCW). We
thank the reviewer for valuable comments.


\end{document}